\documentclass[3p,times]{elsarticle}
\journal{}
\makeatletter
\def\ps@pprintTitle{%
\let\@oddhead\@empty
\let\@evenhead\@empty
\def\@oddfoot{}%
\let\@evenfoot\@oddfoot}
\makeatother
\usepackage{amsmath,amssymb}
\usepackage{tikz}
\usetikzlibrary{arrows.meta}
\usepackage{pgfplots}
\pgfplotsset{compat=1.18}
\usepgfplotslibrary{groupplots}
\usepgfplotslibrary{colorbrewer}

\usepackage{listings}
\usepackage{xcolor}

\lstdefinelanguage{Fortran90}{
  language=Fortran,
  morekeywords={real,integer,implicit,none,allocate,deallocate,call,subroutine,if,then,else,end do,do,optional,save,type,dimension,trim,adjustl,shape},
  sensitive=true,
  morecomment=[l]!,
  morestring=[b]'',
}

\usepackage{listings}
\usepackage{color}

\usepackage{colortbl}
\usepackage{algorithm}
\usepackage{algpseudocode,setspace}
\usepackage{hyperref}
\hypersetup{colorlinks=true,linkcolor=blue,filecolor=magenta,urlcolor=cyan}
\usepackage{xspace}

\usepackage{caption}
\usepackage{subcaption}  
\usepackage{rotating}
\usepackage{graphicx}  
\usepackage{enumitem}
\usepackage[capitalize]{cleveref}
\crefname{lstlisting}{Listing}{Listings}
\Crefname{lstlisting}{Listing}{Listings}

\usepackage{tabularx}
\usepackage{booktabs}
\usepackage{makecell}  

\begin{document}
\captionsetup[table]{skip=2pt} 

\begin{frontmatter} 
\title{SmartFlow: A CFD-solver-agnostic deep reinforcement learning framework for computational fluid dynamics on HPC platforms}

\author[sapienza]{Maochao Xiao} \ead{maochao.xiao@uniroma1.it}
\author[kth]{Yuning Wang}
\author[stuggart]{Felix Rodach}
\author[delft]{Bernat Font}
\author[cwi]{Marius Kurz}
\author[kth]{Pol Suárez}
\author[caltech]{Di Zhou}
\author[kth]{Francisco Alcántara-Ávila}
\author[warwick]{Ting Zhu}
\author[kth]{Junle Liu}
\author[upc]{Ricard Montalà}
\author[uta]{Jiawei Chen}
\author[independent]{Jean Rabault}
\author[bsc]{Oriol Lehmkuhl}
\author[stuggart]{Andrea Beck}
\author[umd]{Johan Larsson} \ead{jola@umd.edu}
\author[kth]{Ricardo Vinuesa} \ead{rvinuesa@mech.kth.se}
\author[sapienza]{Sergio Pirozzoli} \ead{sergio.pirozzoli@uniroma1.it}
\address[sapienza]{Dipartimento di Ingegneria Meccanica e Aerospaziale, Sapienza Università di Roma, Via Eudossiana 18, 00184 Roma, Italy}
\address[kth]{FLOW, Department of Engineering Mechanics, KTH Royal Institute of Technology, Stockholm 10044,
Sweden}
\address[stuggart]{Institute of Aerodynamics and Gas Dynamics, University of Stuttgart, Wankelstraße 3, 70563, Stuttgart, Germany}
\address[delft]{Faculty of Mechanical Engineering, TU Delft, Mekelweg 2, 2628 CD Delft, The Netherlands} 
\address[cwi]{Centrum Wiskunde \& Informatica (CWI), Science Park 123, 1098 XG Amsterdam, The Netherlands}
\address[caltech]{Graduate Aerospace Laboratories, California Institute of Technology, Pasadena, 91125, CA, USA}
\address[warwick]{Department of Statistics, University of Warwick, Coventry, CV4 7AL, UK}
\address[upc]{TUAREG, Universitat Polit`ecnica de Catalunya (UPC), Spain}
\address[uta]{The University of Texas at Arlington, Arlington, TX 76019, USA}
\address[independent]{Independent researcher, Oslo, Norway}
\address[bsc]{Barcelona Supercomputing Center, Barcelona, Spain}
\address[umd]{University of Maryland, College Park, Maryland 20742, USA}

\begin{abstract}

Deep reinforcement learning (DRL) is emerging as a powerful tool for fluid-dynamics research, encompassing active flow control, autonomous navigation, turbulence modeling and discovery of novel numerical schemes. We introduce SmartFlow, a CFD-solver-agnostic framework for both single- and multi-agent DRL algorithms that can easily integrate with MPI-parallel CPU and GPU-accelerated solvers. Built on Relexi and SmartSOD2D, SmartFlow uses the SmartSim infrastructure library and our newly developed SmartRedis-MPI library to enable asynchronous, low-latency, in-memory communication between CFD solvers and Python-based DRL algorithms. SmartFlow leverages PyTorch's Stable-Baselines3 for training, which provides a modular, Gym-like environment API. We demonstrate its versatility via three case studies: single-agent synthetic-jet control for drag reduction in a cylinder flow simulated by the high-order FLEXI solver, multi-agent cylinder wake control using the GPU-accelerated spectral-element code SOD2D, and multi-agent wall-model learning for large-eddy simulation with the finite-difference solver CaLES. SmartFlow's CFD-solver-agnostic design and seamless HPC integration is promising to accelerate RL-driven fluid-mechanics studies.

\end{abstract} 


\end{frontmatter}

\section{Introduction}\label{sec:intro}
Deep reinforcement learning (DRL) has emerged as a versatile and powerful approach in fluid mechanics research, enabling, for instance, the discovery of active control strategies, autonomous navigation, shape optimization, turbulence models and numerical algorithms as summarized in \cref{tab:apps}. Active flow control applications include separation control on flat plates~\citep{font2025deep}, bluff bodies~\citep{suarez2025flow,kurz2025invariant} and airfoils~\citep{garcia2025deep,montala2025deep}, as well as frictional drag reduction~\citep{guastoni2023deep,beneitez2025improving}, and wake stabilization~\citep{yan2023stabilizing}. Beyond control, DRL agents have also been applied to navigate complex fluid environments, such as autonomous urban navigation~\citep{tonti2025navigation} and biological swimming~\citep{koh2025physics}. Shape optimization via DRL has been demonstrated for airfoils~\citep{li2021learning}. In turbulence modeling, DRL has been employed to develop large-eddy-simulation (LES) subgrid-scale closures and wall models for the modeling of unresolved turbulent stresses~\citep{novati2021automating,kurz2025harnessing,heimbach2025reinforcement} and wall shear stress~\citep{bae2022scientific,zhou2024wall,vadrot2023log}. DRL-based methods have also been used to construct zone-adaptive combustion models~\citep{yang2025reinforcement}, RANS closures~\citep{fuchs2024deep}, and nonlinear numerical schemes for conservation laws~\citep{wang2023first,keim2023reinforcement}.

\begin{table}[ht]
    \centering
    \caption{Selected recent deep reinforcement learning applications in fluid mechanics. This table highlights representative state-of-the-art applications and is not intended to be exhaustive.} \label{tab:apps}
    \small
    \begin{tabularx}{\linewidth}{@{}p{0.2\linewidth}X@{}}
    \toprule
    Field of research        & Applications \\
    \midrule
    Active flow control      & Separation control~\citep{font2025deep,suarez2025flow,kurz2025invariant,garcia2025deep,montala2025deep,suarez2025active,yan2025deep,montala2024towards,varela2022deep} \\
        & Frictional drag reduction~\citep{guastoni2023deep,beneitez2025improving,cavallazzi2024deep}\\
        & Wake stabilization~\citep{yan2023stabilizing}\\
        & Suppression of chaotic dynamics~\citep{ozan2025data}\\
        &Rayleigh–Bénard convection control~\citep{vasanth2024multi} \\
    Autonomous navigation  & Navigation in urban flows~\citep{tonti2025navigation}\\
        & Swimming~\citep{koh2025physics,verma2018efficient,gunnarson2021learning,xiong2025chemotactic,qin2023reinforcement} \\
    Shape optimization       & Airfoil shape optimization~\citep{li2021learning,viquerat2021direct} \\
    Turbulence modeling      & LES subgrid-scale and wall models~\citep{novati2021automating,kurz2025harnessing,heimbach2025reinforcement,bae2022scientific,zhou2024wall,vadrot2023log,fischer2025optimal,von2024closure}\\
        & RANS models~\citep{fuchs2024deep}\\
        & Turbulent flame modeling~\citep{yang2025reinforcement} \\
    Numerical algorithms & Nonlinear numerical schemes for conservation laws~\citep{wang2023first,keim2023reinforcement,beck2023toward} \\
    \bottomrule
    \end{tabularx}
\end{table}

Despite the wide range of DRL applications in fluid mechanics, existing DRL frameworks typically exhibit tight coupling to specific CFD solvers and lack full support for deployment on HPC infrastructures, especially GPU‑accelerated supercomputers. A comprehensive overview of frameworks is given in \cref{tab:frameworks}. DRLFEniCS~\citep{rabault2019artificial} uses the FEniCS CPU-based finite element framework to build its CFD solver. FEniCS has a python API, so that the bridging can be performed either directly through native python code importing both the FEniCS and DRL frameworks when running on a single machine or node, or UNIX sockets mechanisms if distributing over several nodes or machines. \citet{guastoni2023deep} used an interface based on the message-passing interface (MPI) to couple the CPU-based solver SIMSON and DRL environment, which are coded in Fortran 77/90 and Python, respectively. The CFD solver is dynamically spawned as a child process, and an intercommunicator is created between the solver and the Python main program. DRLinFluids~\citep{wang2022drlinfluids} integrates the CPU-based OpenFOAM with DRL libraries such as Tensorforce and Tianshou, facilitating active flow control and optimization studies within fluid mechanics. Its adaptable architecture makes DRLinFluids a valuable tool for researchers and engineers. However, the requirement for I/O operations at each control step may introduce a potential bottleneck in training efficiency, though this can be mitigated with the use of, for example, a RAM disk. DRLFluent~\citep{mao2023drlfluent} is a distributed co‑simulation framework that links Python‑based deep reinforcement learning agents with the commercial C++ CFD solver Ansys‑Fluent via a CORBA object‑request broker (omniORB), enabling low‑latency, cross‑node communication on HPC clusters. Its architecture exposes two high‑level interfaces that handle most CFD--RL interactions, reducing the need for labor-intensive secondary coding. DRLFluent integrates seamlessly with Fluent’s closed‑source core; no modifications to the solver itself are required. However, it might require a non-trivial amount of work to replace the CFD solver with a different one.  \citet{novati2021automating} proposed the scientific multi-agent reinforcement learning (SciMARL) framework based on the open-source library Smarties \citep{novati2019a}, which is designed to enable efficient coupling between DRL algorithms and flow solvers. The framework employs the remember and forget experience replay (ReF-ER) method to stabilize learning and reduce the variance in policy updates caused by the simultaneous actions of multiple agents. However, its scalability may be constrained by centralized coordination and the volume of data exchanged between the simulation and learning components, particularly when large flow fields or high-frequency interactions are involved in simulations distributed across many compute nodes. \citet{kurz2022relexi} developed the TF‑Agents‑based Relexi platform. Its advantage is that it offers SmartRedis‑backed scalability, though it is tied to the CPU-based FLEXI CFD solver. Wind-RL~\citep{mole2025reinforcement} employed a similar strategy, using the TorchRL library for reinforcement learning algorithms and the high-order finite-difference Fortran-based Xcompact3d as the backend CFD solver to train active flow controllers for wind turbines. SmartSOD2D~\citep{font2025deep} is developed on top of Relexi and supports GPU-accelerated CFD simulations. However, it remains tightly coupled to the SOD2D CFD solver. 

Beyond these frameworks, several libraries have been proposed to aid DRL-CFD coupling. It is important to distinguish between frameworks and libraries in this context: a framework defines a structured pipeline for DRL-CFD workflow, while a library offers reusable classes or functions that can be integrated into frameworks (pipelines). For example, \citet{shams2023gym} introduced the Gym-preCICE library, which provides a generic adapter class to couple reinforcement learning agents with numerical solvers governed by partial differential equations. The adapter offers a Gymnasium (formerly OpenAI Gym)-like API and uses the preCICE coupling library for backend communication. It supports multiple communication backends, including MPI, TCP/IP, and shared memory. Although flexible, Gym-preCICE requires users to design and implement their own DRL-CFD framework around the adapter, which might involve a non-trivial amount of coding effort. HydroGym~\cite{lagemann2025hydrogym} is a library designed for the application of reinforcement learning to computational flow control research. It offers Gymnasium-compatible environments encompassing different flow-control benchmarks that include both non-differentiable and differentiable formulations. These environments have been evaluated using RL algorithms implemented through different reinforcement learning libraries. HydroGym supports deployment from desktop workstations to high-performance computing clusters. However, HydroGym requires that the target CFD solver expose a Python‑callable interface, which might involve non‑trivial development effort to adapt legacy CFD solvers written in Fortran, C or C++, due to the so-called "two-language gap"~\citep{font2025deep,kurz2022relexi}.

In addition to these code architecture and hardware and software support considerations, several features have emerged as critical to enabling effective use of DRL for flow control. In particular, i) the parallel environment approach (see e.g. \citet{rabault2019accelerating}) and ii) the multi-agent reinforcement learning (MARL, see e.g. \citet{belus2019exploiting}), discussed in more detail in the next section, are key factors to be able to apply DRL to complex flow problems. There too, disparities exist in the level of support offered by different frameworks, and a unified approach would be beneficial.

\begin{table}[ht]
    \centering
    \caption{Open-source DRL frameworks for fluid mechanics. 
    Data Comm.: Data communication between the DRL agent and the CFD solver. 
    Parallel Env.: Parallel CFD environments. 
    GPU: GPU-accelerated CFD solvers.} 
    \label{tab:frameworks}
    \small
    \setlength{\tabcolsep}{4pt}
    \begin{tabularx}{\linewidth}{@{}l l X X X l l@{}}
    \toprule
    Framework                                   & CFD solver            & DRL library                     & Data Comm.                         & Parallel Env. & MARL           & GPU   \\ 
    \midrule
    DRLFEniCS~\citep{rabault2019artificial}     & FEniCS                & TensorForce                     & Native Python / UNIX sockets       & Yes           & No             & No    \\
    Guastoni et al.~\citep{guastoni2023deep}    & SIMSON                & Stable-Baselines3 (PyTorch)     & MPI                                & No            & Yes            & No    \\
    Suárez et al.~\citep{suarez2025flow}        & Alya                  & TensorForce                     & MPI and file I/O stream            & Yes           & Yes            & No    \\
    DRLinFluids~\citep{wang2022drlinfluids}     & OpenFoam              & Tensorforce/Tianshou            & File I/O stream and shared memory  & Yes           & No             & No    \\
    DRLFluent~\citep{mao2023drlfluent}          & Ansys-Fluent          & TensorForce                     & omniORB                            & Yes           & No             & Yes   \\
    sciMARL~\citep{novati2021automating}        & In-house code         & Smarties                        & MPI                                & Yes           & Yes            & No    \\
    Relexi~\citep{kurz2022relexi}               & FLEXI                 & TF-Agents                       & SmartRedis                         & Yes           & Yes            & No    \\
    Wind-RL~\citep{mole2025reinforcement}       & Xcompact3d            & TorchRL                         & SmartRedis                         & Yes           & No             & No    \\
    SmartSOD2D~\citep{font2025deep}             & SOD2D                 & TF-Agents                       & SmartRedis                         & Yes           & Yes            & Yes   \\
    SmartFlow (present)                         & CFD-solver-agnostic   & Stable-Baselines3               & SmartRedis-MPI                     & Yes           & Yes            & Yes   \\
    \bottomrule
    \end{tabularx}
\end{table}

To address these limitations, we have developed SmartFlow, a nearly CFD-solver-agnostic multi-agent DRL framework designed for modern CPU and GPU-accelerated HPC clusters. Here, the term ``multi-agent'' means multiple agents (corresponding to multiple pseudo-environments) with a single shared policy. Built on Relexi~\citep{kurz2022deep} and SmartSOD2D~\citep{font2025deep}, SmartFlow uses the SmartSim~\citep{partee2022using} infrastructure and our newly developed SmartRedis-MPI library to provide asynchronous low-latency in-memory data exchange between Fortran-based CFD solvers and Python-based DRL agents. The newly developed SmartRedis–MPI library makes it trivial to link SmartFlow to a wide variety of CPU or GPU-accelerated CFD solvers. We replace the TensorFlow-based TF-Agents~\citep{guadarrama2018tf} backend with PyTorch’s Stable-Baselines3~\citep{raffin2021stable} for its simplicity and usability, exposing all algorithms through a modular, Gym-inspired~\citep{towers2024gymnasiumstandardinterfacereinforcement} environment API. As \cref{tab:frameworks} shows, SmartFlow can be applied with minimal boilerplate code to single- and multi-agent tasks in the fields of active flow control, autonomous navigation, shape optimization, turbulence modeling, and even numerical scheme discovery. A comparison of SmartFlow with existing frameworks is provided in \cref{tab:frameworks}. SmartFlow employs the above-mentioned two DRL techniques to accelerate training from a computational point of view. The first is to simulate multiple environments in parallel, also known as multi‑environment DRL, so that the agent collects multiple experiences in parallel~\citep{rabault2019accelerating}. The second, orthogonal to the multi‑environment approach, is a multi‑agent reinforcement‑learning method~\citep{belus2019exploiting,vignon2023effective}.

The remainder of this paper is organized as follows. In \cref{sec:rl}, we review the general DRL workflow. In \cref{sec:arch} we detail the SmartFlow architecture and implementation, and in \cref{sec:apps} we present three demonstration cases that showcase the solver-agnostic versatility of SmartFlow.
\begin{enumerate}
    \item Single-agent synthetic-jet control in a cylinder wake using the high-order FLEXI solver
    \item Multi-agent wake control with the GPU-accelerated SOD2D spectral-element code
    \item Multi-agent wall-model learning in LES with the finite-difference CaLES solver
\end{enumerate}
Finally, \cref{sec:conclusion} provides concluding remarks and an outlook on potential future work.

\section{Reinforcement learning}\label{sec:rl}
In this section, the core elements of the classic reinforcement learning paradigm are reviewed. This review is not exhaustive and focuses only on the concepts needed to explain our workflow. For a comprehensive introduction, the reader is referred to~\cite{sutton1998reinforcement}. In reinforcement learning, an agent learns by interacting with an environment over discrete time steps. The interaction is formalized as a Markov decision process (MDP). At each action time step \( t \), the environment occupies a state \( s_t \), and the agent selects an action \( a_t \sim \pi_{\theta}(\cdot \mid s_t) \), where \( \pi_{\theta} \) is the agent policy. This policy maps states to a probability distribution over possible actions, specifying the likelihood of selecting each action given the current state. The selected action \( a_t \) is then applied to the environment, which responds by transitioning to a new state \( s_{t+1} \) according to
\begin{equation}
    s_{t+1} =T\left(s_t, a_t\right),
\end{equation}
where $T$ encodes the underlying system dynamics (e.g., Navier–Stokes equations). The agent simultaneously receives a scalar reward  
\begin{equation}
    r_{t+1} = R(s_t,a_t,s_{t+1}),
\end{equation}
which quantifies performance according to a user‐defined metric. This cycle of observe–act–reward continues until the environment reaches a terminal state $s_T$, producing a trajectory  
\begin{equation}
    \tau = \{(s_0, a_0, r_1),\,(s_1, a_1, r_2),\,\dots,\,(s_{T-1},a_{T-1},r_T)\}.
    \label{eq:trajectory}
\end{equation}
The agent's goal is to learn a policy \( \pi_{\theta} \), which maximizes the expected discounted return, defined from time \( t \) as
\begin{equation}\label{eq:return}
    G_t = \sum_{k=0}^{T-t-1}\gamma^{k}\,r_{t+k+1},
\end{equation}
where the discount factor $\gamma$ balances short‐ and long‐term rewards. This quantity forms the basis of the state value function 
\begin{equation}\label{eq:V}
    V^{\pi}(s_t) = \mathbb{E}_{\pi}\left[G_t \mid s_t\right],
\end{equation}
and state-action value function:
\begin{equation}\label{eq:Q}
    Q^{\pi}(s_t,a_t) = \mathbb{E}_{\pi}\left[G_t \mid s_t, a_t\right].
\end{equation}

\begin{figure}[htbp]
   \centering
   \includegraphics[width=0.835\textwidth,trim = 1.5cm 0.85cm 1.5cm 1.5cm, clip]{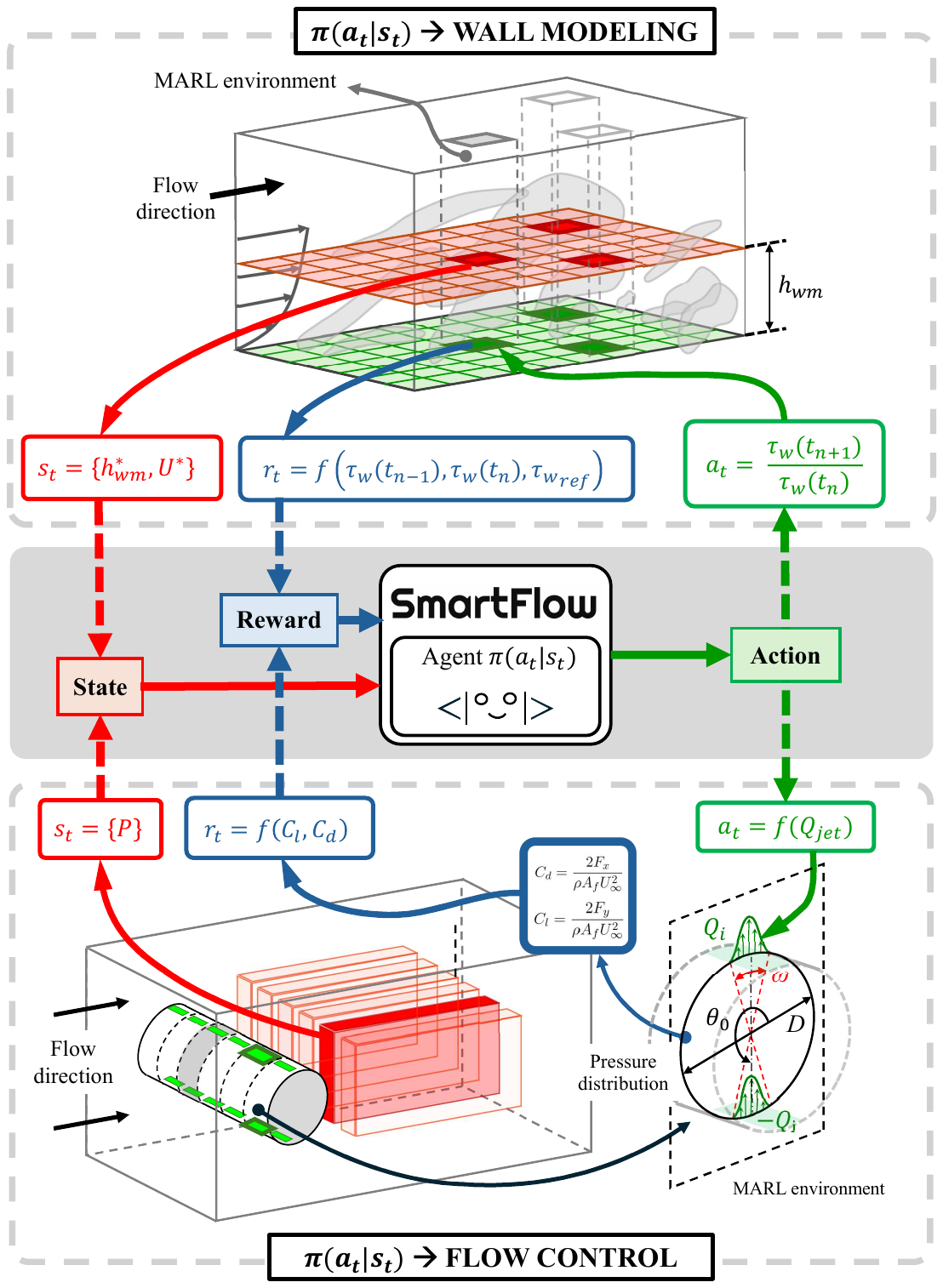}
    \caption{Schematic of the interaction between the SmartFlow framework (middle box) and two distinct CFD environments (top and bottom boxes) within a multi-agent deep reinforcement learning (MARL) setup. 
    Top: In the wall-modeling case, each agent receives a local state \( s_t^{(i)} = \{h^*_{wm}, U^*\} \). The shared policy \( \pi\left(a_t \mid s_t\right) \) selects actions \( a_t^{(i)} = \tau_w(t_{n+1})/\tau_w(t_n) \), and rewards are computed as \( r_t^{(i)} = f(\tau_w(t_{n+1}), \tau_w(t_n), \tau_{w,ref}) \).  
    Bottom: In the flow-control case, agents positioned around a cylinder observe local pressure states \( s_t^{(i)} = \{P\} \) and output control actions \( a_t^{(i)} = f(Q_{jet}) \), modulating the jet mass-flow rate to perturb the wake and reduce drag. Rewards are defined from aerodynamic coefficients, \( r_t^{(i)} = f(C_l, C_d) \), where $C_l$ and $C_d$ are lift and drag coefficients.   
    This illustration is to highlight that the middle-box SmartFlow framework can be easily connected to distinct environments easily. The illustration is inspired by Figure~1 in~\citet{suarez2025active}.
    }
   \label{fig:rl_workflow}
\end{figure}

The SmartFlow training loop adheres to the classic reinforcement‐learning paradigm of environment–agent interaction. \cref{fig:rl_workflow} illustrates this workflow using two example environments, a cylinder flow environment for multi-agent synthetic‐jet control for drag reduction and a turbulent channel flow environment for multi-agent reinforcement learning of LES wall models. In addition to the enforcing of invariants in the control law using MARL, training is accelerated by spawning several CFD simulations in parallel to collect more episodes within a given wall-clock time. At each decision time step \(t\), each agent samples a low‐dimensional observation \(s_t\), such as probe velocities in the cylinder flow control case, and velocity information at a fixed wall-normal height in the wall model case. Based on the observation, the agent (i.e., thanks to MARL, a shared neural-network policy) computes an action \(a_t\), such as jet-blowing/suction rate and wall-stress adjustment, and applies the action to the CFD environment. Then, it receives a scalar reward \(r_t\) that reflects instantaneous performance metrics such as skin‐friction coefficient or drag reduction. During each episode, each environment–agent pair collects a trajectory \(\{(s_t, a_t, r_{t+1})\}_{t=0}^{T-1}\). Once the preset horizon \(T\) is reached, experiences from all parallel instances are gathered to perform a policy update via a reinforcement learning algorithm. Over successive episodes, the shared policy steadily refines its mapping from flow observations to control actions, eventually converging on a strategy that maximizes the expected return as defined by \cref{eq:return}. 

SmartFlow employs two DRL techniques to accelerate training from a computational point of view. The first is to simulate multiple environments in parallel, also known as multi‑environment DRL, so that the agent collects multiple experiences in parallel\citep{rabault2019accelerating}. The second, orthogonal to the multi‑environment approach, is a multi‑agent reinforcement‑learning method. First introduced by \citet{belus2019exploiting} to fluid mechanics and later named as such in the context of flow control by \citet{vignon2023effective}, MARL exploits the spatial invariance of the environment domain to reduce the dimensionality of the action space. For example, consider the channel‑flow environment for wall modeling in the top panel of \cref{fig:rl_workflow}. Because the flow is statistically invariant in the streamwise and spanwise directions, each wall‑surface cell can be treated as its own subdomain (or MARL pseudo‑environment). Instead of predicting multiple actions simultaneously, individual agents, one per pseudo‑environment, each predict a single local action. All agents share the same policy so that each individual learning experience is shared among all agents. This approach reduces the effective action‐space dimensionality and avoids the curse of dimensionality. As reported by~\citet{suarez2025active} and~\citet{vignon2023effective}, using MARL enables the agent to effectively learn the dynamics of the system and a positive control strategy. If we define the number of parallel environments as \(N_e\) and the number of pseudo‑environments per environment as \(N_{pe}\), then \(N_e N_{pe}\) episodes can be sampled in parallel. Consequently, SmartFlow dramatically shortens the data collection time, thus training time. Finally, it is important to highlight the definition of state, action, and reward in the multi‑agent setup within the SmartFlow framework. The state is typically a local state vector, and the action is a single control vector for each pseudo‑environment. However, the reward is flexible: it may be defined locally, globally, or as a combination of both.

The SmartFlow framework relies on model-free reinforcement learning, in which the policy is optimized directly through trial‐and‐error interactions, organized in episodes, rather than model‐based reinforcement learning that relies on learning an explicit model of the dynamics of the environment. Model‐free methods differ in how they gather experience and update the policy, and can be grouped into three broad classes: policy‐gradient, value‐function, and actor‐critic algorithms. Policy-gradient algorithms parameterize the policy \(\pi_\theta\left(a \mid s\right)\) and optimize \(\theta\) to maximize expected return, while value-function algorithms estimate the value functions $V^{\pi}(s_t)$ and $Q^{\pi}(s_t,a_t)$. Actor‐critic algorithms combine both approaches by learning a policy (actor) and a value function (critic) in tandem. Model‐free algorithms can also be on‐policy, using only data collected under the current policy, or off‐policy, which reuse past experiences from a replay buffer.  While off‐policy methods often achieve higher sample efficiency, on‐policy techniques tend to be simpler and more stable. In this work, we employ the clipped proximal policy optimization (PPO) algorithm~\citep{schulman2017proximal}, a model‐free, on‐policy actor‐critic method.  

\section{Software architecture}\label{sec:arch}
SmartFlow is a single- and multi-agent deep reinforcement learning framework designed for a wide array of fluid-mechanics tasks, ranging from active flow control and shape optimization to autonomous navigation in complex flow fields and turbulence modeling. At its core lies the classic RL loop of environment–agent interaction, where agents observe states, take actions, and receive rewards. Built upon Relexi~\citep{kurz2022deep} and SmartSOD2D~\citep{font2025deep}, SmartFlow offers new features and enhancements to improve modularity and performance. \Cref{tab:evol} shows the evolution of the SmartFlow framework from Relexi through SmartSOD2D. Its nearly CFD-solver-agnostic design relies on the SmartRedis-MPI library to bridge Fortran-based CFD codes and Python-based RL agents: any CFD solver (e.g., CaLES, FLEXI, SOD2D) can be easily integrated to the DRL framework. For example, the CaLES solver is integrated into the framework by adding only five lines of code. All reinforcement learning algorithms run in PyTorch via Stable-Baselines3, which provides a cleaner, more flexible API compared to TensorFlow’s TF-Agents. Both agents and CFD solvers can fully leverage GPU acceleration on modern HPC clusters. Defining custom environments in Python requires minimal boilerplate. Training metrics, including rewards, losses, and hyperparameters, are logged automatically to Weights \& Biases or TensorBoard for end-to-end experiment tracking.

\begin{table}[htbp]
\centering
\caption{Evolution of the SmartFlow framework from Relexi through SmartSOD2D.}
\label{tab:evol}
\renewcommand{\arraystretch}{1.2}
\begin{tabularx}{\linewidth}{@{} lXX @{}}
\toprule
Framework & New features / enhancements & Key modifications \\
\midrule
\makecell[tl]{Relexi\\(\citet{kurz2022deep})} & 
\begin{itemize}[leftmargin=*, nosep]
  \item Original scalable reinforcement learning framework
\end{itemize} & 
\begin{itemize}[leftmargin=*, nosep]
  \item SmartSim IL for orchestration on HPC
  \item SmartRedis client for in-memory data communication
  \item TF-Agents as the DRL backend
  \item Coupled to the FLEXI discontinuous Galerkin solver
  \item MPMD supported with OpenMPI via automatically generated rankfiles
\end{itemize} \\
\makecell[tl]{SmartSOD2D\\(\citet{font2025deep})} & 
\begin{itemize}[leftmargin=*, nosep]
  \item GPU-accelerated SOD2D solver support
  \item Environment customization based on a generic CFD-environment base class 
  \item Reduced environment‐startup overhead via parallel environment startup
\end{itemize} & 
\begin{itemize}[leftmargin=*, nosep]
  \item Fork of Relexi
  \item Replace FLEXI with GPU-accelerated SOD2D solver
  \item Customized environment class inherited from a general CFD environment
  \item MPI's MPMD supported by automated allocation of resources on SLURM environments
\end{itemize} \\
\makecell[tl]{SmartFlow\\(present)} & 
\begin{itemize}[leftmargin=*, nosep]
  \item Improved modularity and usability
  \item Nearly CFD-solver-agnostic
  \item Cleaner and more flexible API for reinforcement learning 
  \item Enhanced support for GPU-accelerated CFD solver
  \item Support for multi-task learning
\end{itemize} & 
\begin{itemize}[leftmargin=*, nosep]
  \item Fork of SmartSOD2D
  \item CFD-solver-agnostic architecture by introducing the SmartRedis-MPI library
  \item Switched from TF-Agents to Stable-Baselines3
  \item Enhanced compatibility with GPU-accelerated CFD solvers
  \item Refined \texttt{CFDEnv} interface for easy environment customization
  \item Logging via Weights \& Biases
\end{itemize}\\
\bottomrule
\end{tabularx}
\end{table}

\begin{figure}[htbp]
   \centering
   \includegraphics[width=\linewidth]{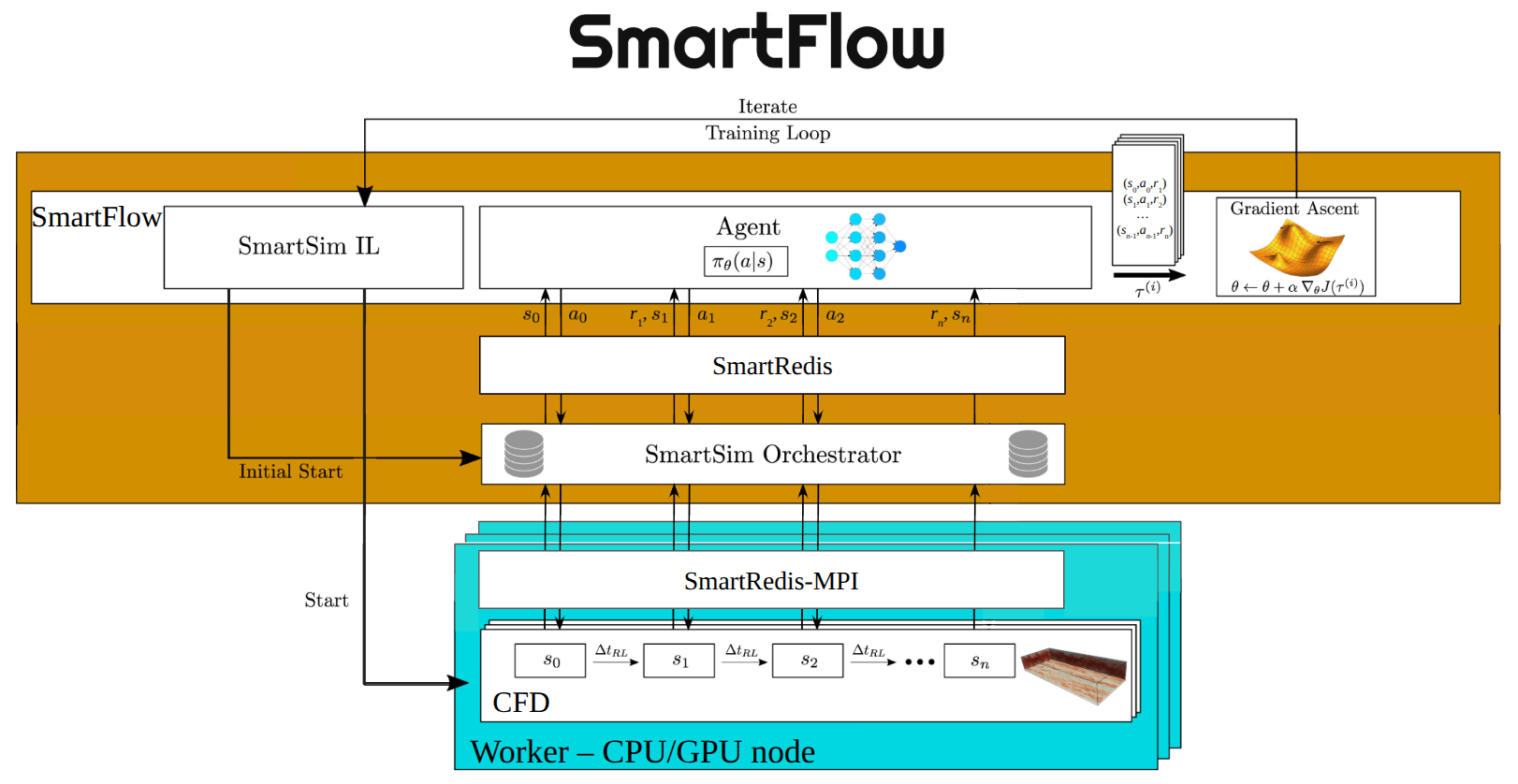}
   \caption{SmartFlow architecture. Before entering the training loop, the SmartSim IL is used to launch the SmartSim orchestrator, which provides an in‐memory database for data exchange between SmartFlow and the HPC workload. At the start of each iteration, the SmartSim IL launches a batch of CFD instances and distributes them to the worker nodes. Each CFD instance could be initialized from a random state and then advanced in time. For every RL update interval $\Delta t_{RL}$, each CFD environment sends its current states to the agents and receives new actions. We apply a synchronous RL algorithm, which means that SmartFlow waits until all simulations have terminated, then aggregates the collected episodes $\tau^{(i)}$ and performs an update on the policy weights $\theta$. Here, $\nabla_{\theta}J\bigl(\tau^{(i)}\bigr)$ denotes the gradient of the expected return evaluated on the current batch of experience. The loop of collect–update is then repeated with the new policy until the policy converges. The figure is adapted from~\citet{kurz2022deep}.}
   \label{fig:arch}
\end{figure}

\Cref{fig:arch} outlines the architecture of SmartFlow as a modular RL framework that implements an RL training loop for fluid-dynamics applications. Coupling HPC application codes with modern machine-learning libraries is often tedious. This is especially true for RL, where the interaction between the HPC application and the RL algorithm is intricate. A considerable part of these difficulties arises from different programming‐language preferences and hardware requirements. Many existing approaches attempt to rewrite machine learning capabilities in HPC languages or re-implement HPC solvers in Python, but this requires rewriting large codebases and does not keep up with ML advances. The parameter update of the policy requires access to the collected trajectories to compute the gradients through backpropagation. At the same time, running CFD simulations requires high MPI parallel efficiency and high GPU speedup rates, in particular, for high-fidelity simulation of real engineering problems, thus compromising either side by code-rewriting is inadvisable. SmartFlow bridges this "two-language gap" by employing the SmartSim infrastructure library to couple a state‐of‐the‐art RL library with modern HPC solvers. Its modular design allows changing RL algorithms and simulation environments with minimal changes to the underlying code. SmartFlow is designed for HPC systems and comprises the following building blocks:

\begin{enumerate}
  \item SmartSim~\citep{partee2022using}: the in‐memory orchestration and communication library,
  \item SmartRedis-MPI: the data communication library developed based on SmartRedis,
  \item CFD solver: any CFD solver for CFD environment simulation,
  \item Stable-Baselines3~\citep{raffin2021stable}: the RL algorithm library.
\end{enumerate}

\subsection{SmartSim}\label{sec:smartsim}
The SmartSim library~\citep{partee2022using} simplifies the coupling of HPC workloads with machine learning workflows. It comprises two core components: the SmartSim infrastructure library (IL) and SmartRedis. The IL handles job orchestration—launching, managing, and distributing MPI‐parallel CFD simulations within a single batch allocation—and can submit jobs automatically to the scheduler. In SmartFlow, we leverage the IL to launch the exact number of CFD environments needed for each rollout, individually or in parallel using MPI's multi-program multi-data (MPMD) mode. While our current implementation opts for individual launches to maximize robustness, future releases will add MPMD support to further reduce CFD-environment‐startup latency.

\subsection{SmartRedis-MPI}\label{sec:smartredis-mpi}
We leverage another key feature of the SmartSim IL: its ability to configure and launch an in-memory, Redis-based database, known as the Orchestrator database, directly on the head node. SmartSim also supports the distribution of the Orchestrator across multiple nodes, but a single instance on the head node is commonly sufficient for our applications. All simulation processes communicate with this Orchestrator via SmartRedis clients (available in Python, Fortran, C, and C++), which support both data exchange and remote procedure calls (e.g., triggering preloaded ML inference routines). The SmartRedis-MPI library is a lightweight library that streamlines data exchange between the parallel CFD solver and Redis database by providing a small set of MPI-aware subroutines written on the basis of SmartRedis. With just a few lines of code, it enables any MPI-parallel CFD application to push and pull data from a Redis database. The SmartRedis-MPI library provides APIs tailored for reinforcement learning tasks. The core APIs in the library are listed in \cref{tab:smartredis-api}. 

\begin{table}[htbp]
\centering
\caption{Key API functions of the SmartRedis-MPI library.}
\label{tab:smartredis-api}
\renewcommand{\arraystretch}{1.2}
\begin{tabularx}{\linewidth}{@{}lX@{}}
\toprule
Function & Description \\
\midrule
\texttt{init\_smartredis\_mpi} & Initialize the MPI environment and creates a SmartRedis client connection on the master process \\
\texttt{put\_state} & Collect the states for all agents across MPI ranks and send them to the Orchestrator datastore \\
\texttt{get\_action} & Retrieve the control actions computed by the RL agent from the Orchestrator datastore and distribute them across MPI ranks \\
\texttt{put\_reward} & Collect the rewards for all agents across MPI ranks and send them to the Orchestrator datastore \\
\texttt{finalize\_smartredis\_mpi} & Cleanly destroy the SmartRedis client connection on the master process \\
\bottomrule
\end{tabularx}
\end{table}

During training, each CFD instance uses SmartRedis-MPI to push flow states into the in-memory Orchestrator. The head-node SmartFlow process then pulls these states via the Python SmartRedis client, computes the next control actions, and writes them back into the Orchestrator. Finally, each CFD instance retrieves its assigned actions through SmartRedis-MPI. The SmartRedis-MPI library is the key to SmartFlow's nearly CFD-solver-agnostic design.

\subsection{CFD solver}\label{sec:cfd_solver}
A CFD solver is typically parallelized via MPI and may use finite‐difference, finite‐volume, spectral‐element, or flux‐reconstruction discretizations. By linking in the SmartRedis-MPI library, often with just a few extra lines of code, each rank can efficiently push its local states and rewards to, and pull control actions from, the shared Orchestrator datastore.

\subsection{Stable-Baselines3}\label{sec:sb3}
SmartFlow builds on the PyTorch‐based Stable‐Baselines3 library for RL algorithms. Custom training environments are created by subclassing the \texttt{VecEnv} API-overriding only \texttt{reset()}, \texttt{step\_async()}, \texttt{step\_wait()}, and a few utility methods—so that all of the CFD coupling details remain encapsulated within the environment class. This design lets any algorithm provided in Stable‐Baselines3 (e.g. PPO, SAC, TD3) interact with the CFD solver. To further streamline development, SmartFlow provides a \texttt{CFDEnv} base class (itself a \texttt{VecEnv}), which implements a generic, asynchronous, vectorized CFD environment; users can simply subclass \texttt{CFDEnv} to implement their own specific CFD environments.

With the four building blocks in place, the SmartFlow workflow can be summarized as follows. \Cref{fig:arch} depicts its modular architecture, and \cref{alg:smartflow} details the multi‐agent RL loop.  Before training, SmartSim IL launches the Orchestrator, an in-memory Redis (or KeyDB) datastore, to exchange data between SmartFlow and the HPC workload.  At each iteration, the IL launches the required CFD instances, each linked via SmartRedis‐MPI.  Every instance pushes its initial states for all its agents to the Orchestrator, which SmartFlow then polls and reads.  The current policy computes actions from these states and writes them back to the datastore. CFD instances poll for their actions, apply them, advance one control interval, and push the next resulting states and rewards. SmartFlow continues to poll, read, and act until the horizon of the episode is reached.  Once a batch of episodes is collected, SmartFlow updates the policy and begins the next collect–update cycle.  This process repeats until the policy converges. Crucially, a single batch of CFD instances can encompass multiple tasks, which might include distinct flow conditions or numerical setups, enabling multi‐task RL; moreover, each task may include several parallel CFD instances within the same batch.

\begin{algorithm}[htbp]
\caption{SmartFlow multi‐agent RL training workflow}\label{alg:smartflow}
\begin{algorithmic}[1]
\State \textbf{Input:} 
  $K$ (number of CFD instances), 
  $N$ (agents per CFD instance), 
  $S_{\max}$ (total environment steps), 
  $T$ (rollout horizon)
\State $S \leftarrow 0$
\State Initialization
\State Launch SmartSim Orchestrator
\While{$S < S_{\max}$}
  \State Clear trajectory buffer $\mathcal{B}$
  \For{$k = 1$ \textbf{to} $K$}
    \State Start CFD instance $k$
  \EndFor
  \For{$k = 1$ \textbf{to} $K$}
    \State Receive initial states $\{s_0^{(k,i)}\}_{i=1}^N$ from CFD instance $k$
  \EndFor
  \For{$t = 0$ \textbf{to} $T-1$}
    \For{$k = 1$ \textbf{to} $K$}
      \State Compute actions $\{a_t^{(k,i)}\}_{i=1}^N$ via policy $\pi_\theta$ from states $\{s_t^{(k,i)}\}$
      \State Put $\{a_t^{(k,i)}\}_{i=1}^N$ to the Orchestrator database 
    \EndFor
    \State Advance all CFD instances by one control time interval $\Delta t_{RL}$
    \For{$k = 1$ \textbf{to} $K$}
      \State Poll and get next states $\{s_{t+1}^{(k,i)}\}_{i=1}^N$ and rewards $\{r_{t+1}^{(k,i)}\}_{i=1}^N$ from the Orchestrator database
    \EndFor
    \State $S \leftarrow S + K \times N$
  \EndFor
  \State Update policy $\pi_\theta$ on buffer $\mathcal{B}$ for a fixed number of epochs
\EndWhile
\State Shutdown SmartSim Orchestrator
\State Finalization
\end{algorithmic}
\end{algorithm}

To further demonstrate the nearly CFD-solver-agnostic nature of SmartFlow and help users easily integrate their own solvers, we provide guidelines for coupling a MPI-parallel CFD solver with the SmartFlow framework. In most cases, only several lines of code are required. As an example, we present the integration of the CaLES solver~\citep{xiao2025cales} with SmartFlow to enable multi-agent deep reinforcement learning of wall models in large-eddy simulation. Listings~\ref{lst:main} and~\ref{lst:wallmodel} show a portion of the main program and the complete DRL-based wall model subroutine (\texttt{wallmodel\_DRL}), respectively. Only five additional lines of code, highlighted in blue, are required for the integration. The 1st and 5th lines---\texttt{init\_smartredis\_mpi} and \texttt{finalize\_smartredis\_mpi}---are inserted into the main program to initialize and finalize the SmartRedis client. The 2nd, 3rd, and 4th lines---\texttt{put\_state}, \texttt{put\_reward}, and \texttt{get\_action}---are added to the \texttt{wallmodel\_DRL} subroutine to communicate with the database. For clarity, GPU-specific code is omitted from Listings~\ref{lst:main} and~\ref{lst:wallmodel}. The integration of the GPU-accelerated CaLES to SmartFlow is available in the CaLES GitHub repository: \url{https://github.com/CaNS-World/CaLES}.

In SmartFlow, a Redis database is used by default to host the Orchestrator. To improve the Orchestrator’s memory efficiency and throughput, the multi‐threaded Redis fork KeyDB can be substituted as the underlying datastore—this has been supported since recent versions of SmartSim and was leveraged in the Relexi framework to improve parallel performance~\citep{kurz2022deep}. A potential bottleneck arises from the overhead of sequentially starting a large number of parallel environments across thousands of MPI ranks. In some scenarios (e.g.\ fast GPU‐accelerated CFD simulations), this startup time can rival or even exceed the simulation runtime. This issue can be addressed by leveraging MPI’s multiple‐program multiple‐data (MPMD) mode, supported by SmartSim, to launch all simulations (each with its own command‐line arguments) in a single MPI call. It also helps to copy required simulation files (parameter and restart files) to each node’s local RAM disk, which could reduce access times compared to a parallel file system like Lustre. In SmartFlow, we dropped mandatory MPMD support to preserve solver‐agnosticism: requiring MPMD would force users to modify their CFD solver to support MPMD before integration. For simplicity, input files remain on the parallel file system. Future SmartFlow releases will offer optional support for these optimizations in performance-critical applications.

\begin{lstlisting}[language=Fortran90, caption={Main program}, label={lst:main}]
call MPI_INIT(ierr)
call MPI_COMM_SIZE(MPI_COMM_WORLD,nprocs,ierr)
call MPI_COMM_RANK(MPI_COMM_WORLD,myid,ierr)
call init_smartredis_mpi(db_clustered, MPI_COMM_WORLD) ! Line 1 of 5 lines of code

...

call finalize_smartredis_mpi ! Line 5 of 5 lines of code
call MPI_FINALIZE(ierr) 
\end{lstlisting}
\begin{lstlisting}[language=Fortran90, caption={DRL-based wall model subroutine}, label={lst:wallmodel}]
subroutine wallmodel_DRL(visc, flattened_state, flattened_stress, flattened_metric)
  implicit none
  real(rp), intent(in) :: visc
  type(FlattenedState), intent(in) :: flattened_state
  type(FlattenedStress), intent(inout) :: flattened_stress
  type(FlattenedMetric), intent(in), optional :: flattened_metric
  real(rp) :: u1, u2, upar, tauw_tot, tauw1, tauw2, this_action
  integer :: n_points, i
  logical, save :: is_first = .true.
  real(rp), allocatable, dimension(:,:), save :: drl_state
  real(rp), allocatable, dimension(:,:), save :: drl_action
  real(rp), allocatable, dimension(:,:), save :: drl_reward
  n_points = size(flattened_state%vel_x1)
  if (is_first) then
    allocate(drl_state (2, n_points))
    allocate(drl_reward(2, n_points))
    allocate(drl_action(1, n_points))
  end if
  drl_state(1, :) = flattened_state%hwm_plus
  drl_state(2, :) = flattened_state%velh_plus
  call put_state(trim(adjustl(tag))//".state", shape(drl_state), drl_state) ! Line 2 of 5 lines of code
  
  if (is_first) then
    is_first = .false.
  else
    drl_reward(1, :) = flattened_metric%tauw1
    drl_reward(2, :) = flattened_metric%tauw1_prev
    call put_reward(trim(adjustl(tag))//".reward", shape(drl_reward), drl_reward) ! Line 3 of 5 lines of code
  end if

  call get_action(trim(adjustl(tag))//".action", shape(drl_action), drl_action) ! Line 4 of 5 lines of code
  flattened_stress%action(:, 1) = drl_action(1, :)
  do i = 1, n_points
    this_action = flattened_stress%action(i, 1)
    tauw_tot = this_action * flattened_stress%tauw(i)
    u1 = flattened_state%vel_x1(i)
    u2 = flattened_state%vel_x2(i)
    upar = sqrt(u1**2 + u2**2)
    tauw1 = tauw_tot * u1 / (upar + eps)
    tauw2 = tauw_tot * u2 / (upar + eps)
    flattened_stress%tauw1(i) = tauw1
    flattened_stress%tauw2(i) = tauw2
  end do
end subroutine wallmodel_DRL
\end{lstlisting}

SmartFlow has been tested across a range of computing environments, from local laptops and workstations to several HPC clusters. These systems include both CPU- and GPU-accelerated configurations, demonstrating the flexibility and portability of the framework. For instance, deployments have been carried out on GPU-enabled partitions such as Booster at CINECA (Italy), N32EA14P at BSCC (China), Alvis (Sweden) and TACC Stampede3 (USA), as well as on CPU-only systems like DCGP at CINECA and BSCC-A at BSCC.

\begin{table}[htbp]
\centering
\caption{Computing systems on which SmartFlow has been tested. NIC refers to the network interface card.}
\label{tab:run_settings}
\small
\renewcommand{\arraystretch}{1.2}
\begin{tabularx}{\linewidth}{@{}llXXll@{}}
\toprule
    Cluster & Partition & Command & Launcher & NIC & GPU acceleration \\
\midrule
    Local laptop/server & / & local / mpirun & local & lo & Yes / No \\
    Single compute node & / & local / mpirun & local & lo & Yes / No \\
    CINECA (Italy) & Booster & srun & Slurm & lo / ib0 & Yes \\
    CINECA (Italy) & DCGP & srun & Slurm & lo / ib0 & No \\
    BSCC (China) & N32EA14P & srun & Slurm & lo / ib0 & Yes \\
    BSCC (China) & BSCC-A & srun & Slurm & lo / ib0 & No \\
    Alvis (Sweden) & / & srun & Slurm & lo & Yes \\ 
    TACC Stampede3 (USA) & skx-dev & srun & Slurm & lo / ib0 & Yes \\ 
\bottomrule
\end{tabularx}
\end{table}

To facilitate the use of the SmartFlow framework for DRL applications in fluid mechanics, we provide a set of usage examples that cover a range of computing architectures. \Cref{tab:usage_examples} provides an overview of the example configurations. All examples focus on the same multi-agent DRL problem for LES wall modeling, as described in~\cref{sec:wallmodel}. Examples~\#1 and~\#2 are designed to run on a single laptop; Examples~\#3 and~\#4 on a standalone local server; Examples~\#5,~\#6, and~\#7 on a single node of an HPC cluster; and Examples~\#8 and~\#9 across multiple nodes (using two nodes in these examples). For each architecture, both CPU-based and GPU-accelerated solvers are considered. All examples use the $Re_\tau = 5200$ turbulent channel flow for training, except for the multi-task example~\#7 on a single node, where multiple Reynolds numbers are used: $Re_{\tau} = u_{\tau} h / \nu = 10^3$, $10^5$, $10^7$, and $10^{10}$.

\begin{table}[htbp]
\centering
\caption{SmartFlow usage examples provided in the SmartFlow GitHub repository. NIC refers to the network interface card, $N_{\text{cases}}$ denotes the number of parallel cases (tasks) for training, $N_{\text{cfds}}$ is the number of CFD simulations per case, and $N_{\text{procs}}$ is the number of MPI processes used per CFD simulation.
}
\label{tab:usage_examples}
\small
\renewcommand{\arraystretch}{1.2}
\resizebox{\textwidth}{!}{%
\begin{tabularx}{\linewidth}{@{}llllllllX@{}}
\toprule
    ID & Example                          & System        & Task type    & Hardware      & Command     & Launcher & NIC & $N_{cases} \times N_{cfds} \times N_{procs}$ \\
\midrule
    1 & laptop\_single\_task\_cpu        & Laptop         & Single-task  & CPU           & mpirun      & local    & lo                & $1 \times 1 \times 1$ \\
    2 & laptop\_single\_task\_gpu        & Laptop         & Single-task  & CPU+GPU       & mpirun      & local    & lo                & $1 \times 1 \times 1$ \\
    3 & local\_server\_single\_task\_cpu & Local server   & Single-task  & CPU           & mpirun      & local    & lo                & $1 \times 4 \times 4$ \\
    4 & local\_server\_single\_task\_gpu & Local server   & Single-task  & CPU+GPU       & mpirun      & local    & lo                & $1 \times 4 \times 1$ \\
    5 & single\_node\_single\_task\_cpu  & Single node    & Single-task  & CPU           & mpirun      & local    & lo                & $1 \times 4 \times 4$ \\
    6 & single\_node\_single\_task\_gpu  & Single node    & Single-task  & CPU+GPU       & mpirun      & local    & lo                & $1 \times 4 \times 1$ \\
    7 & single\_node\_multi\_task\_gpu   & Single node    & Multi-task   & CPU+GPU       & mpirun      & local    & lo                & $4 \times 1 \times 1$ \\
    8 & multi\_nodes\_single\_task\_cpu  & Multiple nodes & Single-task  & CPU           & srun        & slurm    & ib0               & $1 \times 4 \times 4$ \\
    9 & multi\_nodes\_single\_task\_gpu  & Multiple nodes & Single-task  & CPU+GPU       & srun        & slurm    & ib0               & $1 \times 4 \times 1$ \\
\bottomrule
\end{tabularx}
}
\end{table}

%
%

\section{Illustrative applications}\label{sec:apps}
We illustrate SmartFlow’s versatility through three case studies, each coupling a different CFD solver to an RL task in fluid mechanics. Two examples address active flow control: (1) single‐agent drag reduction in 2D cylinder flow via zero-net-mass-flux synthetic jets, with the environments simulated via the high‐order FLEXI solver (discontinuous‐Galerkin spectral element), and (2) multi‐agent drag reduction in 3D cylinder flow using the GPU‐accelerated SOD2D solver (high‐order spectral element). The third example tackles turbulence modeling, using multi‐agent RL to train a wall model for large‐eddy simulation with the GPU‐accelerated CaLES solver (second‐order finite‐difference). In all studies, we employ proximal policy optimization (PPO) with a multilayer‐perceptron (MLP) policy network. These demonstrations underscore SmartFlow’s support for both single‐ and multi‐agent RL across a variety of CFD backends.

\begin{table}[ht]
    \centering
    \caption{Illustrative applications of SmartFlow integrated with different CFD solvers}
    \label{tab:examples}
    \small
    \renewcommand{\arraystretch}{1.2}
    \begin{tabularx}{\linewidth}{@{} l l l l X @{}}
    \toprule
    ID & Purpose               & Number of agents & CFD solver                 & Details \\
    \midrule
    1  & Active flow control   & Single           & FLEXI                      & 2D cylinder flow control through zero-net-mass-flux synthetic jets \\
    2  & Active flow control   & Multiple         & GPU-accelerated SOD2D      & 3D cylinder flow control through zero-net-mass-flux synthetic jets \\
    3  & Turbulence modeling   & Multiple         & GPU-accelerated CaLES      & Wall model development for large-eddy simulation \\
    \bottomrule
    \end{tabularx}
\end{table}

\subsection{Single-agent RL for flow control}
\label{sec:drl_FLEXI}
This demonstration case, following the approach of \citet{rabault2019artificial}, focuses on reducing both the drag and the amplitude of the unsteady lift oscillations acting on a cylinder in laminar flow. Both the drag and the lift fluctuations are significantly increased by vortex shedding behind the cylinder. Therefore, the main control objective is to suppress the vortex shedding. Given the well-characterized periodic nature of this phenomenon, the case represents a relatively simple control problem, making it a well-suited benchmark for RL-based active flow control.

\begin{figure}[htbp]
    \centering
    \begin{tikzpicture}[
    font=\footnotesize
]
    \def\D{1}
    \def\l{8*\D}   
    \def\h{3*\D}  
    \def\cx{2}   
    \def\cy{2}   
    \def\hatch{0.3*\D}  
    \def\r{0.51}  
    \def\umax{0.24} 

    \draw[thick] (-\h/2,-\h/2) rectangle (\l-\h/2,\h/2);
    \draw[|{Stealth}-{Stealth}|,thick] (-\h/2,-\h/2-\hatch) -- (0      ,-\h/2-\hatch) node[midway, anchor=north]{$2D$};
    \draw[|{Stealth}-{Stealth}|,thick] (-\h/2,-\h/2-2.5*\hatch) -- (\l-\h/2,-\h/2-2.5*\hatch) node[midway, anchor=north]{$L=22D$};
    
    \draw[|{Stealth}-{Stealth}|,thick] (-2.5*\hatch-\h/2,-\h/2) -- (-2.5*\hatch-\h/2,\h/2) node[midway, anchor=south, rotate=90]{$H=4.1D$};
    \draw[|{Stealth}-{Stealth}|,thick] (-\h/2-\hatch,-\h/2) -- (-\h/2-\hatch,0) node[midway, anchor=south, rotate=90]{$2D$};
    
    \draw[thick,fill=white] (0, 0) circle (\D/2);
    \draw[{Stealth}-{Stealth},thick] (0,\D/2) -- (0,-\D/2) node[midway, anchor=west]{$D$};

    \draw[domain=70:110, samples=500, variable=\t]
        plot[shift={(0,0)}]
        ({\t}:{0.5 + 0.1*cos(180/40*(\t- 90))}); 
    \draw[domain=250:290, samples=500, variable=\t]
        plot[shift={(0,0)}] 
        ({\t}:{0.5 + 0.1*cos(180/40*(\t-270))}); 

    \foreach \angle in {82, 90, 98} {
        \pgfmathsetmacro{\R}{0.5 + 0.1*cos(180/40*(\angle- 90))}
        \draw[-{Stealth[scale = 0.4]}] (\angle:0.5) -- (\angle:\R);
    }
    \foreach \angle in {262,270,278} {
        \pgfmathsetmacro{\R}{0.5 + 0.1*cos(180/40*(\angle-270))}
        \draw[-{Stealth[scale = 0.4]}] (\angle:0.5) -- (\angle:\R);
    }

  

  \begin{scope}[shift={(-\h/2,0)}]
    \draw[thick, domain=-\h/2+0.01:\h/2-0.01, samples=50, variable=\y]
      plot ({\umax*4*(0.5*\h-\y)*(0.5*\h+\y)/\h^2}, \y);
  \end{scope}

  \foreach \i in {1,...,9} {
      \pgfmathsetmacro\y{-\h/2 + \i*\h/10}      
      \pgfmathsetmacro\u{\umax*4*(0.5*\h - \y)*(0.5*\h + \y)/\h^2}
      \draw[-{Stealth}, thick] (-\h/2 ,\y) -- (\u-\h/2 ,\y); 
    }

  \draw[fill=black] (0, 0.75) circle (0.05);
  \draw[fill=black] (0,-0.75) circle (0.05);

  \draw[fill=black] (0.75, 1) circle (0.05);
  \draw[fill=black] (0.75, 0) circle (0.05);
  \draw[fill=black] (0.75,-1) circle (0.05);

  \draw[fill=black] (1.75, 1) circle (0.05);
  \draw[fill=black] (1.75, 0) circle (0.05);
  \draw[fill=black] (1.75,-1) circle (0.05);

  \draw[fill=black] (2.75, 1) circle (0.05);
  \draw[fill=black] (2.75, 0) circle (0.05);
  \draw[fill=black] (2.75,-1) circle (0.05);
\end{tikzpicture}
    \captionsetup{width=0.8\textwidth}
    \caption{Simulation setup for the flow around a two-dimensional cylinder. Pressure probe locations are indicated by black circles.}
    \label{fig:DC1_setup}
\end{figure}
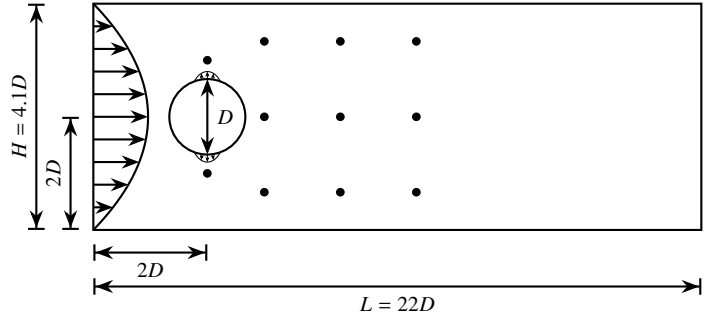

Based on the ``2D-2'' benchmark case proposed by~\citet{schafer1996benchmark}, the two-dimensional simulation environment features a circular cylinder of diameter $D$, placed within a channel of size $L\times H = 22D\times 4.1D$, where $L$ and $H$ are the sizes in the streamwise and wall-normal directions, respectively. The cylinder axis is vertically offset by $0.05D$ from the channel centerline to promote vortex shedding. A schematic of this configuration is shown in \cref{fig:DC1_setup}. At a Reynolds number $Re_D = U_bD/\nu = 100$ and a Mach number of $Ma = U_b/\sqrt{\gamma RT} = 0.2$, both defined with respect to $D$ and the bulk velocity $U_b$, the flow behaves quasi-incompressible. This ensures consistency with the reference results of \citet{rabault2019artificial}, despite the use of the compressible solver FLEXI~\citep{krais_flexi_2021}.

The control employs two synthetic jets, each following a cosinusoidal velocity distribution along the cylinder's angular coordinate $\varphi$. The profile is symmetrically confined to an angular width $\omega$ centered at $\varphi_i = \pm\pi/2$, such that
\begin{equation}
    u_{\perp,i}(\varphi) = 
    \begin{cases}
        Q_i \frac{\pi}{2\omega D^2}\cos{\left(\frac{\pi}{\omega}(\varphi-\varphi_i)\right)},   & \left|\varphi - \varphi_i\right|\leq \frac{\omega}{2} \\
        0,      & \left|\varphi - \varphi_i\right|  >  \frac{\omega}{2}
    \end{cases}.
\end{equation}
Zero total mass flow rate is enforced by requiring $\sum_i^n Q_i = 0$, where $Q_i$ denotes the mass flow rate of the $i$-th of $n$ jets. For the present $n=2$ case, this constraint reduces the control input to a single scalar quantity $Q=Q_1=-Q_2$. The nondimensional control input is then defined as $Q^* = Q/Q_{D}$, where $Q_D$ is the reference mass flow rate, computed as the integral of the inflow mass flux over the cylinder's projection onto the inflow boundary. A comprehensive description of the simulation setup is provided in \citet{kurz2025invariant}, with additional technical details discussed by \citet{rabault2019artificial}.

In this simulation environment, we search for an optimal control strategy using single-agent reinforcement learning. Both the policy and the value function are represented by MLPs with two hidden layers of 64 neurons each, employing a hyperbolic tangent activation function. Since MLPs inherently rely on the input ordering, they do not require explicit positional encoding in the state representation. The state consists of relative pressure values measured at 11 probes distributed near the jets and within the cylinder wake, as illustrated in \cref{fig:DC1_setup}. Based on this state, a scalar action is sampled from a Gaussian distribution parameterized by the policy network, reflecting the control input $Q^*$. The sampling is restricted to a bounded action space of $|Q^*| \leq 0.068$ to prevent non-physical control inputs and numerical instabilities near the actuation region. For training, a modified version of the reward function of~\citet{rabault2019artificial} is used, which is based on the ratio of the achieved to the expected maximum drag reduction, rather than the drag itself. This normalizes the drag reduction values to be within approximately $[0,1]$, and also increases the sensitivity of the reward signal to drag changes, thereby facilitating the convergence of training. Building on the approach introduced by~\citet{cavallazzi2024deep}, the reward is defined as
\begin{equation}
    r_t = \dfrac{\langle c_D \rangle^{UC} - \langle c_D \rangle}{\langle c_D \rangle^{UC} - \langle c_D \rangle^{EC}} - \alpha\left|\langle c_L \rangle\right|\;,
\end{equation}
\noindent where $c_D$ and $c_L$ are the drag and lift coefficients, respectively, with respect to the reference length $D$ and the bulk velocity $U_b$. The operator $\langle\cdot\rangle$ represents the exponential temporal moving average, emphasizing the most recent values. Superscripts $UC$ and $EC$ refer to the uncontrolled reference case and the expected minimum under effective control, respectively.

The training was performed using a full batch comprising 26 CFD environments in parallel, each generating a trajectory of 80 steps per episode. The simulations were initialized from randomly selected initial fields drawn from a predefined set of four, and then proceeded for 20 non-dimensional time units,  $t^*=tU_b/D$. The complete training consisted of 400 episodes and required approximately $4.6$ hours on 52 AMD EPYC 7543 cores. After that, the learned policy reached sufficient convergence for demonstration purposes.

\begin{figure}[htbp]
    \centering
    \begin{tikzpicture}[font=\footnotesize]
    \begin{groupplot}[
        cycle list/Dark2,
        group style={
            group name=plot,
            group size=2 by 1,
            yticklabels at=edge left,
            horizontal sep=2pt
        },
        scale only axis,
        width=0.4\linewidth,
        height=0.15\linewidth,
        grid=both,
        grid style={line width=.1pt, draw=gray!5},
        major grid style={line width=.2pt,draw=gray!25},
        ymin = -5, ymax = 35
    ]
    
    \nextgroupplot[
        xmin = 0, xmax =400,
        xlabel=Episodes,
        xlabel style={ at=(ticklabel cs:0.625) },
        ylabel=Average return,
        ylabel near ticks,
    ]
    \addplot+[thick] table [
        x=episode,
        y=return,
        col sep=comma
    ]{democase1_avg_return_ref.csv};
    \addplot+[thick] table [
        x=episode,
        y=return,
        col sep=comma
    ]{democase1_avg_return.csv};
    \nextgroupplot[
        xmin = 1350, xmax =1450,
        xtick={1400},
        width=0.1\linewidth,
        yticklabels=\empty,
        xlabel={},
        ylabel={},
    ]
    \addplot+[thick] table [
        x=episode,
        y=return,
        col sep=comma
    ]{democase1_avg_return_ref.csv};
    \end{groupplot}
\end{tikzpicture}
    \captionsetup{width=0.8\textwidth}
    \caption{Evolution of the average training return for the MLP policies of \citet{kurz2025invariant} (\protect\tikz[baseline=-0.55ex] \protect\draw[color=Dark2-A,thick] (0,0) -- (0.3,0);)
    and the present study (\protect\tikz[baseline=-0.55ex] \protect\draw[color=Dark2-B,thick] (0,0) -- (0.3,0);). The $x$-axis is truncated to skip intermediate episodes. The plot resumes at episode 1350 to show the converged values of the reference.}
    \label{fig:DC1_avg_return}
\end{figure}
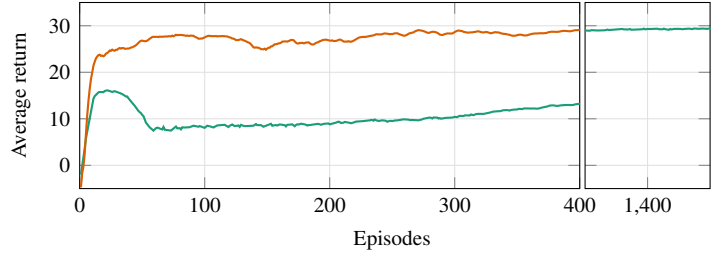

The evolution of the averaged return is shown in \cref{fig:DC1_avg_return}, which demonstrates fast convergence. The pronounced fluctuations in the average return indicate that the policy has not yet fully converged. The fluctuation may also be influenced by the elevated learning rate, which is set five times higher than in prior work of~\citet{kurz2025invariant}. Despite different convergence rates, the final average returns are comparable. These differences in convergence speed are attributed to variations in the PPO implementations and parameter setups.

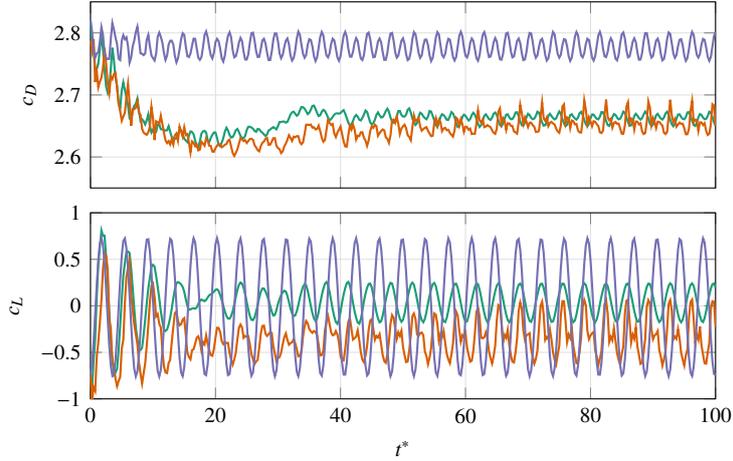
\begin{figure}[htbp]
    \centering
    \begin{tikzpicture}[font=\footnotesize]
    \begin{groupplot}[
        cycle list/Dark2,
        group style={
          group size=1 by 2,
          vertical sep={0.02\textwidth}
        },
        ylabel near ticks,
        width=0.5\linewidth,
        height=0.15\linewidth,
        grid=both,
        grid style={line width=.1pt, draw=gray!5},
        scale only axis, 
        major grid style={line width=.2pt,draw=gray!25},
        xmin = 0, xmax =100
    ]

  \nextgroupplot[
    ymin = 2.55, ymax = 2.85,
    xticklabels={,,},
    ylabel={$c_D$}
  ]

    \addplot+[thick] table [
        x=Time,
        y=cd,
        col sep=comma
    ]{data_ref.csv};
    \addplot+[thick] table [
        x=Time,
        y=cd,
        col sep=comma
    ]{data.csv};
    \addplot+[thick] table [
        x=Time,
        y=cd,
        col sep=comma
    ]{data_uc.csv};
    
    \nextgroupplot[
    ymin = -1, ymax = 1,
    xlabel=$t^*$,
    ylabel={$c_L$},
  ]
    
    \addplot+[thick] table [
        x=Time,
        y=cl,
        col sep=comma
    ]{data_ref.csv};
    \addplot+[thick] table [
        x=Time,
        y=cl,
        col sep=comma
    ]{data.csv};
    \addplot+[thick] table [
        x=Time,
        y=cl,
        col sep=comma
    ]{data_uc.csv};
    \end{groupplot}
\end{tikzpicture}
    \captionsetup{width=0.8\textwidth}
    \caption{Long-term evolution of the drag ($c_D$) and lift ($c_L$) coefficients, comparing the uncontrolled case (\protect\tikz[baseline=-0.55ex] \protect\draw[color=Dark2-C,thick] (0,0) -- (0.3,0);) to the MLP results from \citet{kurz2025invariant} (\protect\tikz[baseline=-0.55ex] \protect\draw[color=Dark2-A,thick] (0,0) -- (0.3,0);) and the present study (\protect\tikz[baseline=-0.55ex] \protect\draw[color=Dark2-B,thick] (0,0) -- (0.3,0);).}
    \label{fig:DC1_longterm}
\end{figure}

To evaluate the long-term behavior of the learned policy, a greedy evaluation run was conducted for a time period up to $t^*=100$. The results, shown in \cref{fig:DC1_longterm}, demonstrate clear improvements over the uncontrolled case in both objectives: drag reduction and mitigation of lift oscillations. However, the reference policy exhibits a notably different behavior. While our MLP generally surpasses the reference in drag reduction, it induces a pronounced shift toward negative lift coefficients. This could likely be reduced, if desirable, by increasing the lift penalization coefficient $\alpha$ in the reward function, as highlighted by \citet{rabault2019artificial}.

\subsection{Multi-agent RL for flow control of bluff body}
\label{sec:drl_SOD2D}

This case illustrates the flow-control application of SmartFlow integrated with the spectral-element SOD2D solver. The target is to use DRL for active flow control (AFC) to reduce drag in bluff bodies, which is of great interest in many engineering applications. While single-agent DRL has shown great potential in simple AFC applications, especially for 2D flows~\citep{vignon2023effective}, it faces limitations due to the curse of dimensionality when dealing with 3D flows exhibiting complex dynamics~\citep{belus2019exploiting}. The multi-agent reinforcement learning mitigates this challenge by dividing a domain into smaller subdomains (i.e., pseudo-environments), making high-dimensional AFC problems more tractable. This setup enables policies to exploit spatial invariance and optimize local rewards, thereby enhancing exploration efficiency.

We consider a 3D cylinder at $Re = 100$ ($Re = U_{\infty} D / \nu$, where $U_{\infty}$ is the inflow velocity, $D$ is the cylinder diameter, and $\nu$ is the kinematic viscosity). This setup has been investigated by \citet{suarez2025flow} and \citet{montala2024towards}. 
The flow is governed by the incompressible Navier--Stokes equations. The computational domain spans $L_x = 30D$, $L_y = 15D$, and $L_z = 4D$ in the streamwise, cross-stream, and spanwise directions, respectively. The cylinder is placed at $(x,y) = (7.5D, 7.5D)$ and extends the full length $L_z$ in the spanwise direction. At the inlet, a constant free-stream velocity $U_{\infty}$ is prescribed, while zero-gradient conditions with constant pressure are applied at the outlet. Slip conditions are enforced on the top and bottom surfaces, and periodic boundary conditions are applied in the spanwise direction.


\Cref{fig:rl_workflow} illustrates the MARL configuration applied to the cylinder. The AFC is achieved through ten pairs of wall-mounted actuators ($n_{\rm jet} = 10$) which are uniformly distributed along the spanwise direction of the cylinder. Each pair consists of one actuator on the top surface ($\theta_0 = 90^{\circ}$) and one on the bottom ($\theta_0 = 270^{\circ}$). The actuators have a spanwise width of $0.4D$ and an angular width of $\omega = 10^{\circ}$ in the $x-y$plane. To ensure instantaneous mass conservation, each actuator pair operates under a zero-net-mass-flux (ZNMF) condition by imposing opposite mass flow rates: $Q_{\rm top} = -Q_{\rm bot}$. Each actuator pair defines one pseudo-environment, resulting in ten pseudo-environments in total. The action of the DRL agent is the mass flow rate $Q$, which determines the imposed velocity at the wall via the following Dirichlet boundary condition:

\begin{equation}
    ||U_{\rm jet}(Q,\theta)|| = Q \frac{\pi}{\rho D \omega}\cos\left( \frac{\pi}{\omega} (\theta - \theta_0) \right),
    \label{eq:sod_cyl_action}
\end{equation}
\noindent where $Q = \dot{m} / L_z$ and $|\theta - \theta_0|\in [ -{\omega}/{2}, {\omega}/{2} ]$, $\dot{m}$ is the mass flow rate. The absolute value of the jet velocity is projected into the $x$ and $y$ axes, since $||U_{\rm jet}||$ corresponds to the radial cylinder direction. To balance the learning efficiency and numerical stability, $|Q| \leq 0.176$ was selected based on the prior study~\citep{suarez2025flow}.

Each pseudo-environment includes $85$ witness points located around the cylinder wall at its local mid-span ($z$-middle) position. Due to the periodic boundary condition in the spanwise direction, the state of each pseudo-environment also includes its two neighboring ones, resulting in $3 \times 85 = 255$ points per state vector.

The lift and drag coefficients are defined as $C_l = l / (1/2 \rho U^2_{\infty}S)$ and $C_d = d / (1/2 \rho U^2_{\infty}S)$, respectively, where $l$ and $d$ denote the lift and drag forces, respectively. 
The reference area $S$ is defined based on the cylinder diameter $D$ and the spanwise length of the pseudo-environment. Taking $C_l$ and $C_d$ into account, the reward of a pseudo-environment $R_i$ is defined as 
\begin{align}
    r_i &= \left ( C_{d, {\rm ref}} - C_d \right) - \alpha | C_l | \\
    R_i &= (1-\beta) r_i + \frac{\beta}{n_{\rm jets}} \sum^{n_{\rm jets}}_{j=1} r_j,      
\label{eq:sod_cyl_reward}
\end{align}
\noindent where $C_{d, {\rm ref}}$ denotes the reference $C_d$, and $\alpha$ and $\beta$ are weighting parameters. In particular, $\alpha$ weights the penalty of $C_l$, while $\beta$ balances the contributions of local and global rewards. We use $\alpha = 0.3$ and $\beta = 0.2$ in the present study.

The PPO algorithm is used for training the DRL agent, with both actor and critic modeled as multi-layer perceptrons comprising two hidden layers of 512 neurons each. Each episode lasts $T_{\rm episode} \approx 34.88\ D/U_{\infty}$, equivalent to approximately six vortex shedding cycles, during which $120$ actions are applied ($T_{\rm action} = T_{\rm episode} / 120$). We trained the agent for $100$ episodes. To improve sampling efficiency, four CFD simulations are performed in parallel. Since each simulation contains $10$ pseudo-environments, $40$ trajectories are collected in total per policy update, resulting in a batch size of $40$. Hyperparameters are chosen based on~\citep{suarez2025flow,montala2024towards} to reproduce the results. However, those studies used TF-Agents or Tensorforce, while the present work adopts Stable-Baselines3. Since PPO implementations can differ across libraries, some variation in performance can be expected. However, detailed hyperparameter tuning is beyond the scope of this demonstration, whose primary goal is to illustrate the functionality of the SmartFlow framework.

\begin{figure}[ht]
    \centering
    \begin{subfigure}{\linewidth}
        \centering
        \includegraphics[width=0.5\linewidth]{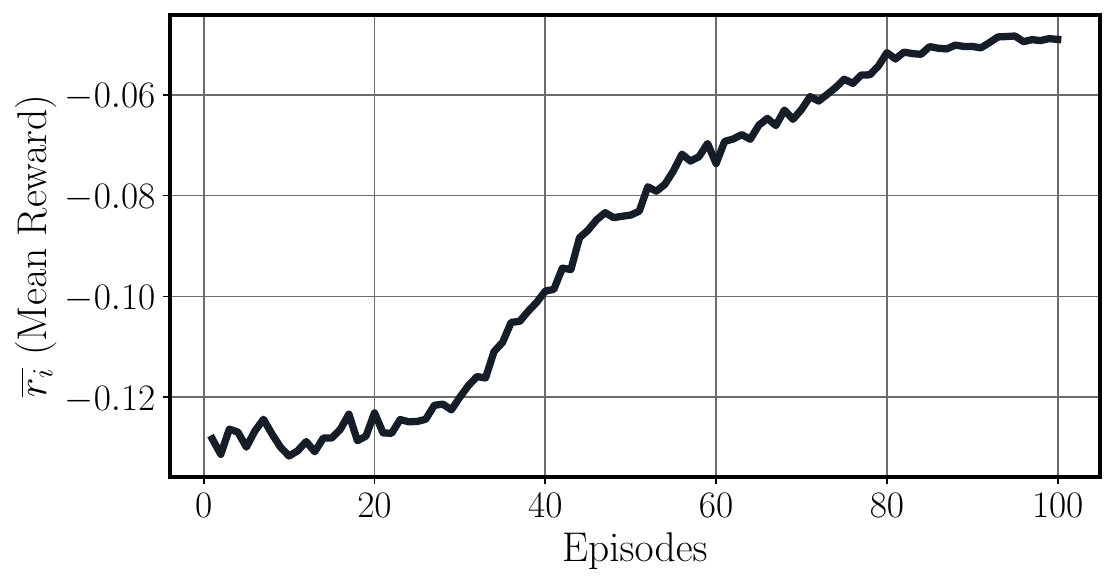}
        \subcaption{Training curve of policy network during exploration.}
        \label{fig:train_sod_cyl}
    \end{subfigure}
    \quad
    \begin{subfigure}{\linewidth}
        \centering
       \includegraphics[width=0.5\linewidth]{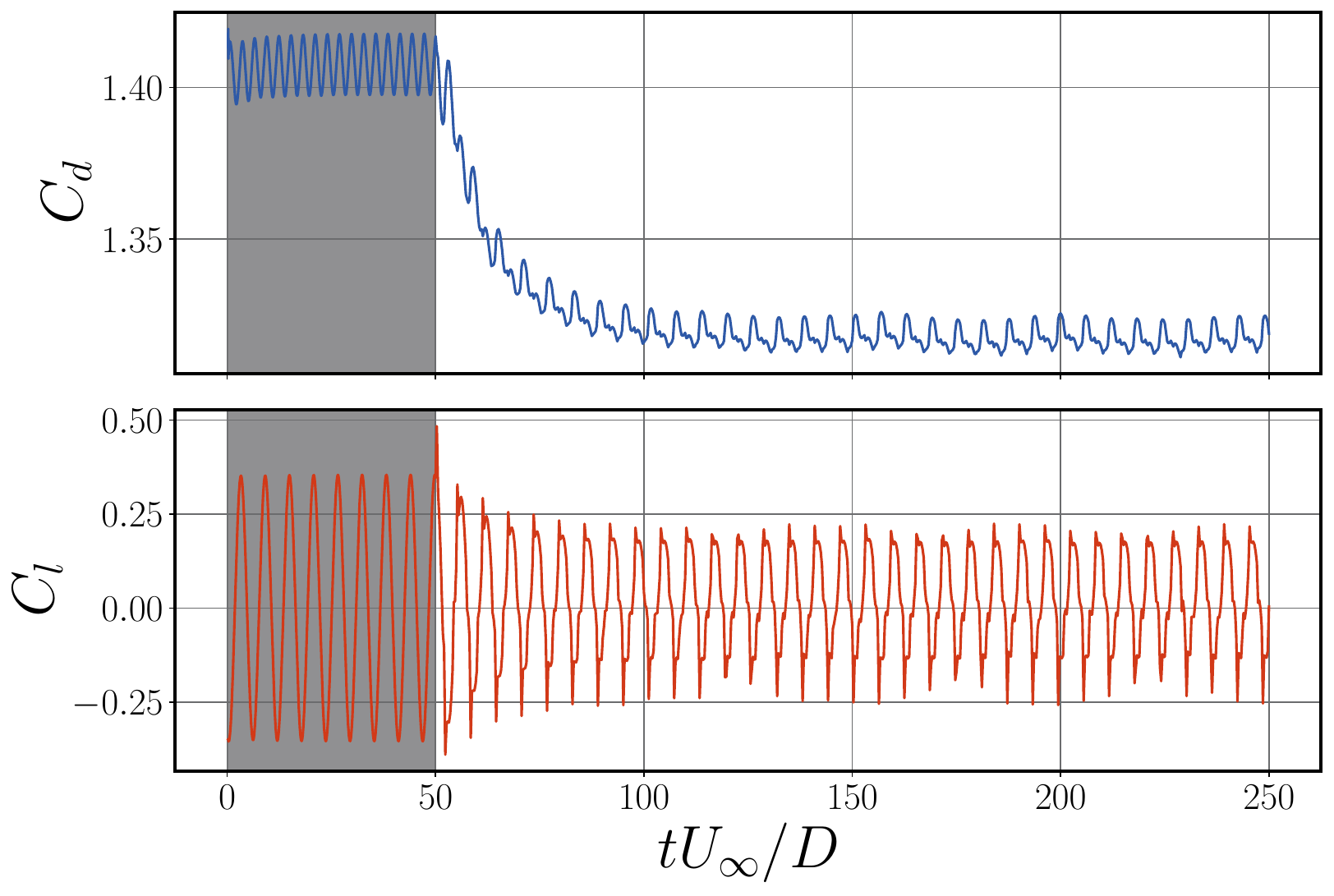}
        \subcaption{Temporal evolution of $C_l$ and $C_d$ before and after activating DRL control in deterministic mode.}
       \label{fig:eval_sod_cyl}
        \end{subfigure}
    \caption{(a) Training curve of the policy networks for drag-reduction control using SOD2D. Note that the $\overline{r}_i$ denotes the mean reward of all trajectories for a single episode. 
00    (b) Temporal evolution of the drag ($C_d$) and lift coefficient ($C_l$) before and after applying DRL control in deterministic mode, where the gray regions denote the uncontrolled evolution of $C_d$ and $C_l$. Note that $t U_{\infty} / D$ is the dimensionless time unit.}
    \label{fig:sod_cyl}
\end{figure}

The mean reward history is shown in~\Cref{fig:train_sod_cyl}. It shows that the curve converges to approximately $-0.04$ in the final 20 episodes. This value is slightly lower than that in~\citet{montala2024towards}, likely due to differences in PPO implementations. The training was performed on a single node with four NVIDIA A100 GPUs and required approximately $40$ hours for $100$ episodes.

The learned policy is evaluated in deterministic mode. \Cref{fig:eval_sod_cyl} presents the time evolution of $C_d$ and $C_l$ before and after DRL activation. Initially, an uncontrolled simulation is run for $50 D/U_{\infty}$ to establish baseline values (gray regions). Once the control is activated, significant reductions in both $C_d$ and $C_l$ are observed. The control is maintained for an additional $200D/U_{\infty}$ to ensure convergence. Relative to the baseline, the mean drag coefficient is reduced by $6.0\%$, and the lift coefficient fluctuation (i.e., standard deviation) by $46.6\%$, demonstrating the effectiveness of MARL for 3D AFC applications.

\subsection{Multi-agent RL for turbulence modeling} \label{sec:wallmodel}
The demonstration case illustrates the application of SmartFlow, integrated with the CaLES solver~\citep{xiao2025cales}, to train a wall model for large‐eddy simulation via multi‐agent reinforcement learning. The wall model is trained in turbulent channel flows at $Re_{\tau} = u_{\tau}h/\nu = 10^3$, $10^5$, $10^7$ and $10^{10}$, where $h$ is the channel half‐height, $u_\tau$ the friction velocity, and $\nu$ the kinematic viscosity. Rather than a single Reynolds number, we include four Reynolds numbers during training, hoping the learned model can generalize across a wide range of $Re_\tau$. At each rollout, four CFD runs, one per Reynolds number, are executed in parallel (additional instances per $Re_\tau$ can be launched if desired). This is the so-called multi-task learning. Each wall-modeled LES (WMLES) run is initialized with the velocity field obtained from the same Reynolds-number WMLES simulation with the equilibrium log-law wall model, with its initial wall‐shear stress randomized between $80\%$ and $120\%$ of the true value. The simulations are driven by constant pressure gradients, so the mean wall shear stress is known \emph{a priori}. For each CFD environment, the matching location $h_{wm}$ is randomly selected between $0.075h$ and $0.150h$ to ensure training over a smooth range of inner-scaled matching locations within the log layer. The velocity is then interpolated to the chosen wall‐normal location $h_{wm}$ to form the state vector. Agents are arranged with spacings $\Delta x_{RL}=4\Delta x$ and $\Delta z_{RL}=4\Delta z$, so there are 192 agents per CFD environment. Each training iteration advances the CFD simulations for 120 action time steps ($120\Delta t_{RL}$, with $\Delta t_{RL}u_b/h\approx0.4$). Prior to agent interaction, each CFD run undergoes a warm‐up of $\Delta tu_b/h=96$ with fixed wall‐shear stress (initalized wall-stress values) to remove the initial transients. The domain size is $L_x=6.4h$, $L_y=2.0h$, $L_z=2.4h$ with $\Delta x=\Delta z=0.1h$, and the wall‐normal mesh stretches from $\Delta y_w=0.025h$ at the wall to $\Delta y_c=0.100h$ at the center.  

\begin{figure}[htbp]
   \centering
   \includegraphics[width=0.8\linewidth]{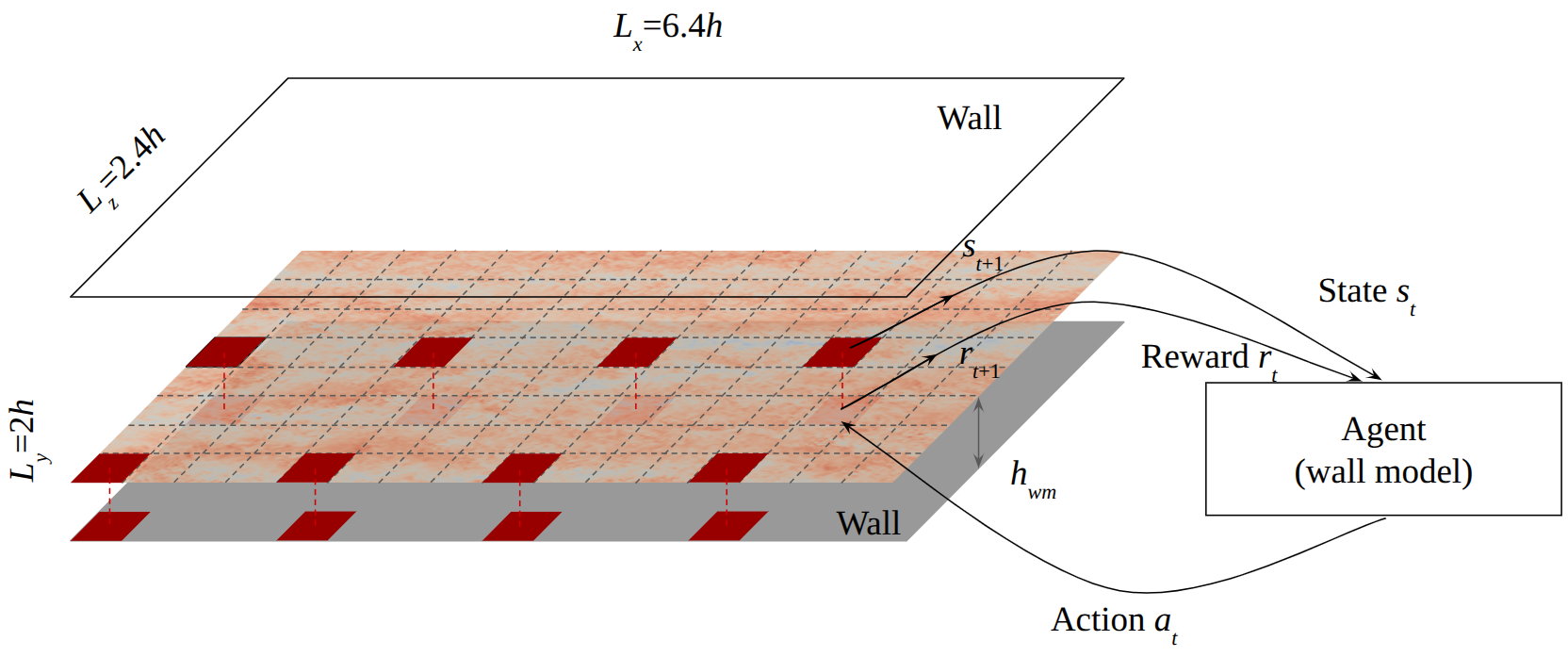}
   \caption{Channel environment setup. Agents are distributed evenly along the wall; each agent obtains state information at a wall‐normal height $h_{wm}$, computes the reward at the wall, and provides the state to the policy $\pi_\theta$, which then outputs the action for the next time step. Dashed lines denote grid lines. The figure is inspired by the illustration in~\citet{bae2022scientific}.}
   \label{fig:env_wm}
\end{figure}

The policy is represented by a multilayer perceptron with two hidden layers of 64 units each and Tanh activations. At each time step, the network outputs the mean \(\mu_\theta\) and log‐standard deviation \(\log\sigma_\theta\) of a diagonal Gaussian, from which continuous actions are sampled. We train the network using the Adam optimizer with a fixed learning rate of \(3\times10^{-4}\), a minibatch size of 92\,160 transitions (i.e.\ the full batch), and 10 gradient‐descent epochs per policy update. Each action from the policy network is interpreted as a multiplicative factor \(a_n(x,z)\in[0.9,1.1]\) on the current wall-shear stress:
\begin{equation}\label{eq:tauw}
\tau_w(x,z,t_{n+1}) = a_n(x,z)\,\tau_w(x,z,t_n).
\end{equation}
The reward is defined as
\begin{equation}
    r(x, z, t_{n+1}) = -\left( \frac{\left| \tau_w^{ref} - \tau_w(x, z, t_{n+1}) \right| - \left| \tau_w^{ref} - \tau_w(x, z, t_n) \right|}{\tau_w^{ref}} \right) + Bonus, 
    \label{eq:1}
\end{equation}
\noindent where
\begin{equation}
    Bonus = \mathbb{I}\left( \frac{\left| \tau_w^{ref} - \tau_w(x, z, t_{n+1}) \right|}{\tau_w^{ref}} < 0.01 \right),
\end{equation}
\noindent where $\mathbb{I}$ is an indicator function and $\tau^{ref}_w$ is the true mean wall‐shear stress. This gives a reward that is proportional to the improvement in the prediction of the wall‐shear stress compared to the one obtained in the previous time step with an additional reward if the predicted wall‐shear stress is within $1\%$ of the true value.  These action and reward definitions follow \citep{bae2022scientific}.  Our state vector definition differs from \citep{bae2022scientific}: rather than using log-law slope and intercept computed from the instantaneous LES flow information at time $t_n$, we take the logarithmic wall‐normal height \(\ln(h_{wm}^*)\) of the matching point and the instantaneous wall-parallel velocity magnitude \(\,U^*(h_{wm})\), where the superscript \(^*\) denotes normalization by the wall units at time \(t_n\). The state vector is
\begin{equation}
  s_t = (h^*_{wm},U^*_{wm}).
\end{equation}
The filtered incompressible Navier–Stokes equations are solved by CaLES using wall‐modeled LES with a staggered, second‐order finite‐difference spatial discretization, a fractional‐step projection method, and a third‐order low‐storage Runge–Kutta time integrator.  Subgrid‐scale stresses are closed with the classical Smagorinsky model, with model coefficient $C_s=0.11$, augmented by a van Driest damping function near the wall.  For the channel flows, periodic boundary conditions are imposed in the streamwise and spanwise directions, while wall-stress boundary conditions with no‐penetration are imposed at the top and bottom walls.  The wall‐shear stress \(\tau_w\) computed via Eq.~\eqref{eq:tauw} is first evaluated at wall‐cell centers; its components \(\bigl(\tau_{w,x},\tau_{w,z}\bigr)\) are then obtained by assuming the stress vector aligns with the wall‐parallel velocity at the matching point.  These components are interpolated to the midpoints of wall‐cell edges and enforced as Neumann boundary conditions on the LES domain:
\begin{equation}
    \tau_{w,x} = \nu \frac{\partial u}{\partial y}, \tau_{w,z} = \nu \frac{\partial w}{\partial y},
\end{equation}
where the subscript \(w\) denotes evaluation at the wall.  Further details on CaLES’s wall-modeled LES implementation and validation are provided in~\citep{xiao2025cales}.  

The training campaign comprised 800 CFD runs, yielding a total of 153\,600 agent episodes (192 episodes per CFD instance).  All simulations were carried out on a single node of the CINECA Leonardo Booster partition, where each node features an Intel Xeon Platinum 8358 CPU (2.60 GHz, 32 cores) and four NVIDIA A100 GPUs (64 GB HBM2e) interconnected by NVLink 3.0 (200 GB/s intra‐node bandwidth). Each batch of four concurrent CFD simulations completed in approximately 12 seconds, and the full training required 40 minutes.  

\begin{figure}[htbp]
   \centering
   \includegraphics[width=0.6\linewidth]{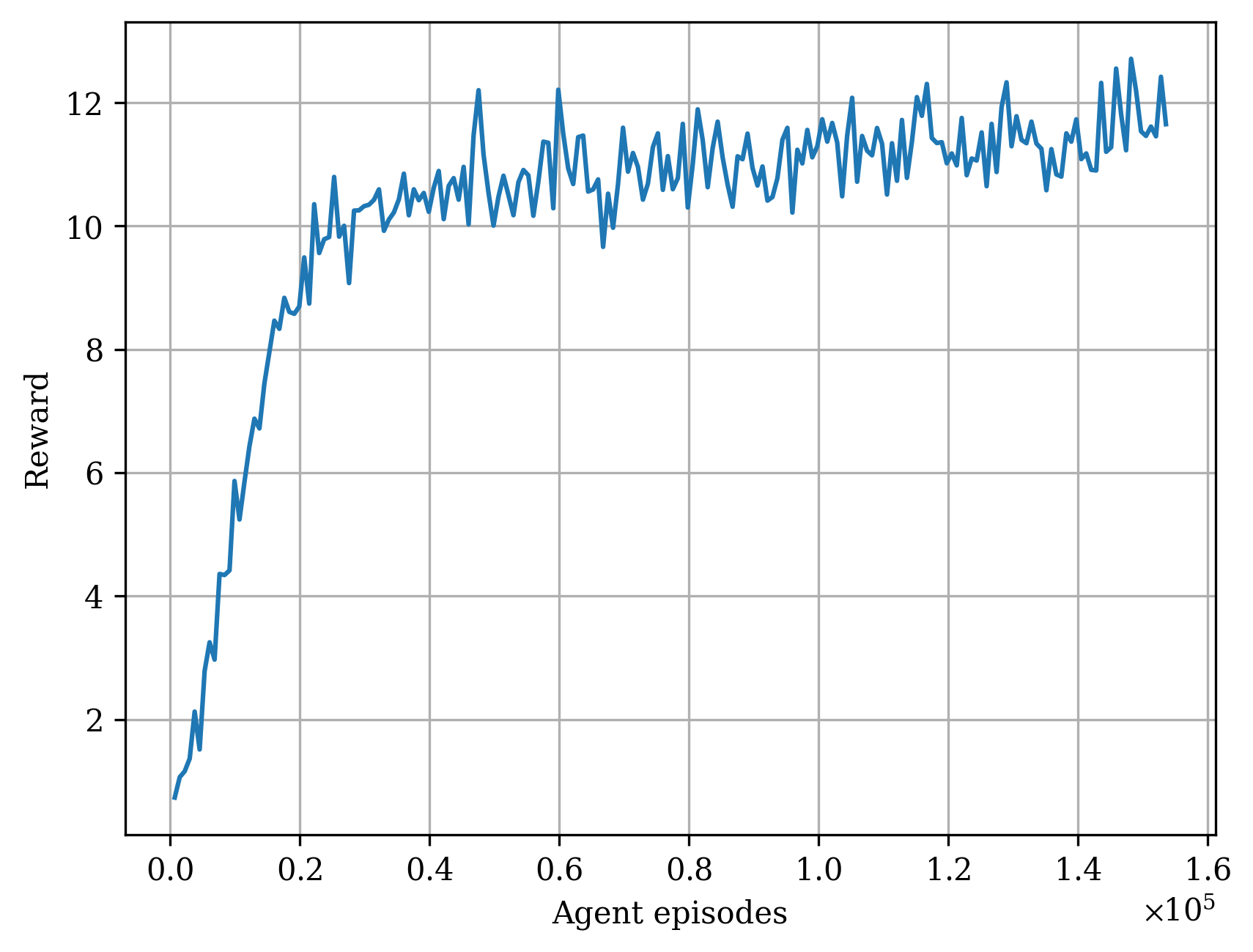}
   \caption{Evolution of the agent’s episode reward versus agent episode number. The reward is averaged over one batch of CFD simulations; the values are raw and no moving average is applied.}
   \label{fig:logging}
\end{figure}

\begin{figure}[htbp]
   \centering
   \includegraphics[width=0.6\linewidth]{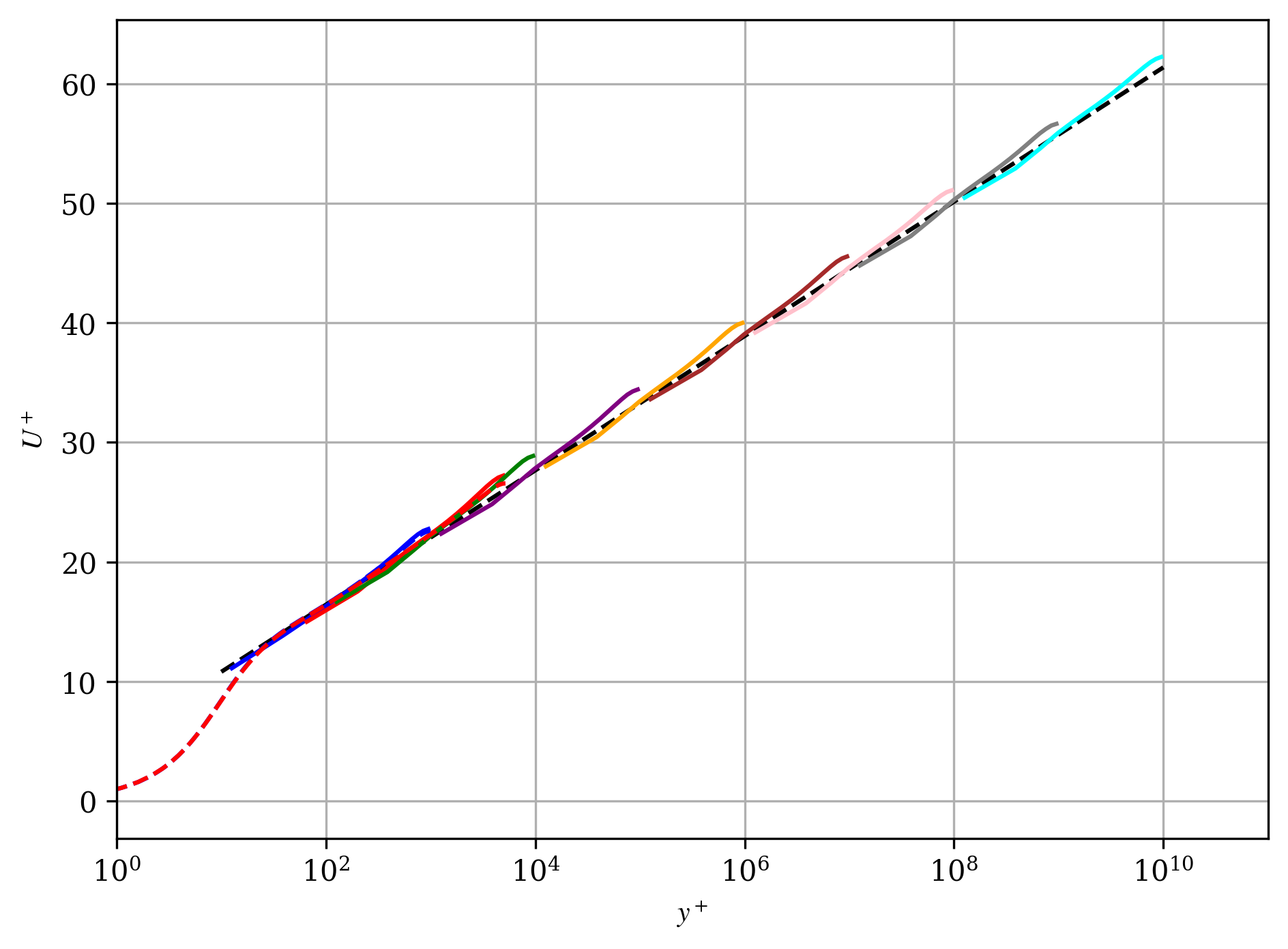}
   \caption{Mean streamwise velocity $U^+$ as a function of the wall‐normal distance $y^+$, where the superscript $+$ indicates normalization by mean wall units. The different colors from bottom left to top right denote $Re_\tau = 10^3$, $5200$, $10^4$, $10^5$, $10^6$, $10^7$, $10^8$, $10^9$, $10^{10}$. Reference DNS results at $Re_\tau = 1000$ (blue dashed) and $5200$ (red dashed)~\citep{lee2015direct}. The log law (black dashed) corresponds to $u^+ = 1/\kappa \ln(y^+) + B$, with $\kappa = 0.41$ and $B = 5.2$.}
   \label{fig:loglaw}
\end{figure}

\Cref{fig:logging} plots the agent’s episode reward averaged over each batch of four simulations. Convergence is effectively reached after about 76\,800 agent episodes (i.e.\ 400 CFD runs), although we completed all 800 runs to guarantee full convergence. We then evaluated the final wall model (trained over 800 runs). In fact, its performance is nearly identical to the model obtained after 400 runs.  \Cref{fig:loglaw} presents the mean velocity profiles from evaluation runs driven by a constant pressure gradient, with the wall‐model matching height fixed at \(h_{wm}=0.1h\).  Nine Reynolds numbers were tested, \(Re_\tau = 10^3,\;5200,\;10^4,\;10^5,\;10^6,\;10^7,\;10^8,\;10^9,\;\text{and}\;10^{10}\).  In all cases, LES coupled with the trained wall model accurately reproduces the reference profiles up to \(Re_\tau=10^{10}\).

\section{Conclusions}\label{sec:conclusion}
Deep reinforcement learning is rapidly growing into a powerful general-purpose toolbox for fluid dynamics, from active flow control and autonomous navigation to shape optimization, turbulence modeling and the discovery of new numerical algorithms. In this work, we introduced SmartFlow, a single‐ and multi‐agent RL framework designed to operate seamlessly on modern HPC clusters and to interface with a wide variety of CFD solvers with minimal boilerplate. Built on top of Relexi and SmartSOD2D, SmartFlow leverages the SmartSim infrastructure library to orchestrate CFD simulations and the SmartRedis client for in‐memory data exchange via the Orchestrator datastore. Key features include an in‐memory SmartRedis‐MPI data‐transfer layer between the CFD solver and the Orchestrator datastore, full support for GPU‐accelerated CFD solvers, and a modular Gym‐style environment interface that can be subclassed for environment customization in Python. Crucially, SmartFlow is effectively CFD‐solver‐agnostic, making it straightforward to integrate with any CPU‐ or GPU‐accelerated CFD code.

To demonstrate the solver‐agnostic design and versatility of SmartFlow, we integrated three distinct CFD backends and trained RL agents on three fluid-mechanics problems. First, we applied a single‐agent policy to zero-net-mass-flux synthetic‐jet control in cylinder flow using the high‐order FLEXI solver. Second, we extended this setup to a 3D-flow multi‐agent case with the GPU‐accelerated SOD2D spectral‐element solver. Third, we developed a multi‐agent wall model for large‐eddy simulation by coupling SmartFlow to the finite‐difference CaLES solver to learn accurate wall‐shear predictions across a wide range of Reynolds numbers. In each case, SmartFlow’s in‐memory orchestration and asynchronous data exchange incurred minimal overhead.

SmartFlow thus offers a nearly solver‐agnostic framework for deploying both single‐ and multi‐agent RL in fluid mechanics on modern GPU-based HPC infrastructure, lowering the barrier to entry for researchers seeking to leverage reinforcement learning as a tool. In future work, we will further enhance SmartFlow and the SmartRedis‐MPI library to improve data‐exchange efficiency in large‐scale RL training.

\section*{Acknowledgements}
We gratefully acknowledge the CINECA award under the ISCRA initiative and the EuroHPC Joint Undertaking for providing access to the LEONARDO high-performance computing resources. Ricardo Vinuesa acknowledges financial support from the ERC grant no. 2021-CoG-101043998, DEEPCONTROL. Felix Rodach and Andrea Beck are grateful for the support by the Friedrich und Elisabeth Boysen-Stiftung BOY-187, the European High Performance Computing Joint Undertaking (JU) under grant agreement
No 101093393 and the state of Baden-Württemberg under the project Aerospace 2050 MWK32-  7531-49/13/7 ``FLUTTER''/``QUASAR''. Marius Kurz carried out this work during the tenure of an ERCIM ``Alain Bensoussan'' Fellowship Programme. We thank H. Jane Bae (California Institute of Technology) for her valuable feedback on the wall model training case and for sharing her implementation that couples the Smarties reinforcement learning library with an in-house large-eddy simulation code. We are grateful to Andrew Mole (Imperial College London) for providing and reviewing the information regarding the Wind-RL framework, Yiqian Mao and Shan Zhong (University of Manchester) for the DRLFluent framework, Ahmed H. Elsheikh (Heriot-Watt University) for the Gym-preCICE library, and Christian Lagemann (University of Washington) for the Hydrogym library.

\section*{Data availability}
The SmartFlow framework is publicly available at \url{https://github.com/SmartFlow-AI4CFD/SmartFlow}. Its primary dependencies are included as Git submodules to ensure consistent versioning:
\begin{enumerate}
    \item SmartRedis‑MPI: \url{https://github.com/SmartFlow-AI4CFD/smartredis-mpi}
    \item SmartRedis: \url{https://github.com/CrayLabs/SmartRedis}
    \item GPU-accelerated CaLES solver: \url{https://github.com/CaNS-World/CaLES}
\end{enumerate}
Example integrations demonstrating the CFD-solver‑agnostic design of SmartFlow:
\begin{enumerate}  
    \item SmartFlow + GPU-accelerated SOD2D (multi‑agent cylinder wake control): \\ \url{https://github.com/SmartFlow-AI4CFD/SmartFlow-SOD2D}
    \item SmartFlow + FLEXI (single‑agent cylinder flow control): \\
    This may be available upon reasonable request, as the specific branch of the FLEXI CFD solver has not been open-sourced.
\end{enumerate}

\bibliography{main_bibfile}
\end{document}